\documentclass{an}
\usepackage{graphicx}
\usepackage{times}

\begin{document}

\Pagespan{1}{}
\Yearpublication{2010}
\Yearsubmission{2010}
\Month{1}
\Volume{1}
\Issue{1}
\DOI{DOI}
\title{Optical multiband surface photometry of a sample of Seyfert galaxies. \\III. Global, isophotal, and bar parameters\thanks{Based on observations obtained with the 2-m telescope of the Institute of Astronomy and National Astronomical Observatory, Bulgarian Academy of Sciences.}}
\author{L. Slavcheva-Mihova\inst{}\thanks{Corresponding author:
\email{lslav@astro.bas.bg}} \and B. Mihov\inst{}}
\titlerunning{Optical multiband surface photometry of Seyfert galaxies}
\authorrunning{L. Slavcheva-Mihova \& B. Mihov} 
\institute{Institute of Astronomy and National Astronomical Observatory, Bulgarian Academy of Sciences, \\
72 Tsarigradsko Chaussee Blvd., 1784 Sofia, Bulgaria}
\received{R-date} \accepted{A-date}
\publonline{P-date}
\abstract{This paper is third in a series, studying the optical properties of a sample of Seyfert galaxies.
Here we present a homogeneous set of global (ellipticity, position angle, inclination, and total magnitude) and isophotal (semi-major axis and colour indices at 24 $V$ mag arcsec$^{-2}$) parameters of the galaxy sample. 
We find the following median corrected isophotal colour indices: 
 $(B-I_{\rm \scriptstyle C})_{24}^{(0)}$$\,=\,$$1.9$\,mag\,arcsec$^{-2}$  and
 $(V-I_{\rm \scriptstyle C})_{24}^{(0)}$$\,=\,$$1.1$\,mag\,arcsec$^{-2}$.
A set of bar parameters~-- ellipticity, position angle, semi-major axis corresponding to the ellipticity maximum in the bar region, and length, are also reported; deprojection has been applied to the bar ellipticity, length, and relative length in terms of galaxy isophotal semi-major axis.
Regarding bar length estimation, we use a method, based on the relation between the behaviour of the profiles and orbit analysis.
The so estimated bar length tightly correlates with the semi-major axis, corresponding to the ellipticity maximum with a median ratio of the former to the latter of 1.22. 
The median of the deprojected bar ellipticity, length, and relative length are 0.39, 5.44\,kpc, and 0.44, respectively. 
There is a correlation between the deprojected bar length and the corrected isophotal semi-major axis at 24 $V$ mag arcsec$^{-2}$. 
Three of the 17 large-scale bars appear strong, based on the deprojected bar ellipticity as a first-order approximation of bar strength. 
The deprojected relative bar length does not appear to correlate with the bar ellipticity.
 }
\maketitle

\section{Introduction}

The major components of disk galaxies are bulges and disks, basically different in their support against gravitational collapse (\cite{J3_96}).
Various correlations involving bulge and disk parameter have been established, e.g., bulge vs. disk scale lengths (MacArthur,~Courteau~\& Holtzman \cite{MCH_03}; \cite{M_04}; \cite{AEC_05}), bulge effective surface brightness (SB) vs. Hubble type (\cite{J3_96};\linebreak \cite{M_04}), bulge effective colour index (CI) vs. disk central CI (\cite{J4_96}).  
Furthermore, bulge-to-disk ratio underlies morphological classification of disk galaxies.
Correlations between bulge parameters and black hole mass (\cite{FF_05}) evidence the coevolution of the\linebreak black hole and its host galaxy.
Considering active galaxies, the problems related to the origin and angular momentum reduction mechanisms of the fuel have given rise to much discussion (e.g., Jogee \cite{J_06}).
In this regard, comparative analysis of the morphology and local environment of matched active and inactive galaxy samples have been performed (e.g., \cite{MR_97}; De~Robertis, Yee \& Hayhoe \cite{RYH2_98}; Virani, De~Robertis \& VanDalfsen \cite{VRV_00};\linebreak \cite{SSF_07}).
Thus, studies on the fueling me\-chanisms of active galactic nuclei and correlations among galaxy parameters, as well as the precise morphological classification, all demand morphological characterization, i.e., disclosure of the features present. 
We analysed the evidence of non-axisymmetric perturbation of the potential in a sample of 35 Seyfert galaxies and in a matched sample of inactive galaxies, based on a detailed morphological characterization and study of the local environment.
The results are presented in Slavcheva-Mihova \& Mihov (\cite{SM_11}, hereafter Paper\,I). 
Here we present a homogeneous set of global, isophotal, and bar parameters  of the  Seyfert galaxies\footnote{Three more galaxies~-- Mrk\,1040, NGC\,5506, and Mrk\,507, were added.}.

Global and isophotal parameters are involved in inclination corrections and various galactic structure studies. 
While databases provide global and isophotal parameters for the bulk of the galaxies, they are generally based on photographic data of less photometric accuracy than CCD data.

The classical bar signature on the profiles is an ellipticity maximum, accompanied by a position angle (PA) plateau and a SB bump (Wozniak \& Pierce \cite{WP_91}; Wozniak et~al. \cite{WFM_95}); more detailed bar criteria were introduced later on (Knapen, Shlosman \& Peletier \cite{KSP2_00}; Men\'endez-Delmestre et~al. \cite{MSS_07}; Marinova \& Jogee \cite{MJ_07}; Aguerri, M\'endez-Abreu \& Corsini \cite{AMC_09}). 
There are various methods for bar length estimation,
based on: visual inspection of images, analysis of
the SB profile along the bar major axis, isophote fitting with ellipses, Fourier analysis, etc. (see the reviews of Athanassoula \& Misiriotis \cite{AM_02}; Erwin \cite{E_05}; Michel-Dansac \& Wozniak \cite{MW_06}). 
The semi-major axis (SMA) corresponding to the ellipticity maximum ($\ell_{\rm max}$, see Wozniak \& Pierce \cite{WP_91}) proved to be the most robust, objective, and reproducible among the bar length estimates. 
It, however, underestimates bar length (Wozniak et~al. \cite{WFM_95}) and is not related to any of the bar dynamical characteristics (e.g., Michel-Dansac \& Wozniak \cite{MW_06}).
Moreover, bar strength can be defined as the maximum tangential force in terms of the mean radial force after \cite{CS_81}. Thus, it generally depends on the bar ellipticity, bar mass, and central force field. 
The tight correlation found between bar strength and deprojected bar ellipticity shows that the latter is a good measure of bar strength (Laurikainen, Salo \& Rautiainen \cite{LSR_02}). 
Values of the deprojected ellipticities below 0.15 are among the signatures of ovals and lenses (\cite{KK_04}). 
 
 Details about the sample selection, observations, data reduction, and Johnson-Cousins $BVR_{\rm \scriptstyle C}I_{\rm \scriptstyle C}$ surface photometry, as well as contour maps and profiles of the SB, CI, ellipticity ($\epsilon$), and PA could be found in Paper\,I.
The extra added galaxies were reduced as the rest of sample; observational details, contour maps, and profiles are presented in Appendix\,\ref{AppendixA}. 

The paper is structured as follows.
In Sect.~\ref{glopar} we present the global values of the ellipticity, PA, inclination, and total magnitude.
The isophotal parameters~-- SMA and CIs at 24 $V$ mag arcsec$^{-2}$, are discussed in Sect.~\ref{isopar}.
The bar parameters are outlined in Sect.~\ref{barpar}.
 A summary of our results is given in Sect.\,\ref{summary}.
 A set of contour maps and profiles of the 
extra added galaxies is presented in Appendix\,\ref{AppendixA}. 
Global ellipticities and deprojected bar ellipticities of the matched inactive galaxies are given in Appendix\,\ref{AppendixB}.

Throughout the paper the linear sizes in kpc have been calculated using the cosmology-corrected scale given in \linebreak NED\footnote{NASA/IPAC Extragalactic Database.} ($H_0$$=$73 km s$^{-1}$ Mpc$^{-1}$, $\Omega_{\,\rm M}$$=$0.27, $\Omega_{\,\Lambda}$$=$0.73,\linebreak Spergel et~al. \cite{SBD_07}).

\begin{table}[ht]
 \caption{Global values of the ellipticity, PA, and inclination, estimated over all available passbands.}
\label{T_glopar}
\begin{center}
 \begin{tabular}{@{}l@{\hspace{0.2cm}}l@{\hspace{0.2cm}}l@{\hspace{0.3cm}}r@{\hspace{0.3cm}}l@{}}
\hline
\noalign{\smallskip}
~~Galaxy     & Region$^{\rm a}$     & ~~Ellipticity     & PA~~~~~~~    & Inclination    \\
           & (arcsec)     &                 & (degree)~~~           & ~~(degree)         \\
\noalign{\smallskip}\hline
\noalign{\smallskip}
         
Mrk\,335   & $ ~~10      $ & $0.09 \pm 0.02$ & $109.2 \pm 13.7$ & $25.0 \pm 3.0$ \\
III\,Zw\,2 & $ ~~12      $ & $0.17 \pm 0.02$ & $ 14.7 \pm~~4.7$ & $34.7 \pm 2.2$ \\
Mrk\,348   & $ ~~20\,(45)$ & $0.09 \pm 0.02$ & $ 90.1 \pm 41.3$ & $25.0 \pm 3.0$ \\
I\,Zw\,1   & $ ~~12      $ & $0.11 \pm 0.02$ & $139.0 \pm 10.9$ & $27.7 \pm 2.7$ \\
Mrk\,352   & $ ~~~~7     $ & $0.21 \pm 0.01$ & $ 84.6 \pm~~1.1$ & $38.7 \pm 1.0$ \\
Mrk\,573   & $ ~~30      $ & $0.13 \pm 0.02$ & $ 76.2 \pm~~6.8$ & $30.2 \pm 2.5$ \\
Mrk\,590   & $ ~~15      $ & $0.06 \pm 0.02$ & $127.4 \pm 14.5$ & $20.4 \pm 3.8$ \\
Mrk\,1040  & $ ~~20      $ & $0.73 \pm 0.03$ & $ 74.2 \pm~~1.1$ & $79.3 \pm 2.8$ \\
Mrk\,595   & $ ~~16      $ & $0.32 \pm 0.01$ & $ 92.3 \pm~~1.2$ & $48.4 \pm 0.8$ \\
3C\,120	  & $ ~~24      $ & $0.27 \pm 0.02$ & $115.0 \pm~~1.5$ & $44.2 \pm 1.8$ \\
Ark\,120   & $ ~~10      $ & $0.13 \pm 0.01$ & $  8.5 \pm~~2.3$ & $30.2 \pm 1.2$ \\
Mrk\,376   & $ ~~11      $ & $0.29 \pm 0.02$ & $164.8 \pm~~2.5$ & $45.9 \pm 1.8$ \\
Mrk\,79	  & $ ~~47      $ & $0.14 \pm 0.02$ & $148.6 \pm~~2.5$ & $31.4 \pm 2.4$ \\
Mrk\,382   & $ ~~17      $ & $0.16 \pm 0.02$ & $155.1 \pm~~6.8$ & $33.6 \pm 2.3$ \\
NGC\,3227  & $ ~~80      $ & $0.47 \pm 0.01$ & $150.1 \pm~~3.1$ & $59.9 \pm 0.8$ \\
NGC\,3516  & $ ~~35      $ & $0.19 \pm 0.01$ & $ 44.3 \pm~~4.5$ & $36.8 \pm 1.0$ \\
NGC\,4051  & $ 110       $ & $0.39 \pm 0.03$ & $109.3 \pm~~6.3$ & $54.0 \pm 2.4$ \\
NGC\,4151$^{\rm b}$&$~\ldots$&$0.07 \pm 0.03$& $ 26.0 \pm~~3.5$ & $21.0 \pm 5.0$ \\
Mrk\,766   & $ ~~24      $ & $0.19 \pm 0.03$ & $ 66.9 \pm~~4.1$ & $36.8 \pm 3.2$ \\
Mrk\,771   & $ ~~11      $ & $0.13 \pm 0.01$ & $ 80.2 \pm 11.7$ & $30.2 \pm 1.2$ \\
NGC\,4593  & $ ~~75      $ & $0.35 \pm 0.02$ & $ 70.6 \pm~~6.5$ & $50.9 \pm 1.6$ \\
Mrk\,279   & $ ~~18      $ & $0.33 \pm 0.01$ & $ 30.8 \pm~~2.3$ & $49.3 \pm 0.8$ \\
NGC\,5506  & $ ~~40      $ & $0.75 \pm 0.03$ & $ 87.8 \pm~~0.7$ & $81.2 \pm 2.8$ \\
NGC\,5548  & $ ~~35\,(60)$ & $0.19 \pm 0.01$ & $ 96.2 \pm~~6.0$ & $36.8 \pm 1.0$ \\
Ark\,479   & $ ~~10      $ & $0.31 \pm 0.02$ & $117.2 \pm~~1.5$ & $47.6 \pm 1.7$ \\
Mrk\,506   & $ ~~12      $ & $0.27 \pm 0.01$ & $116.3 \pm~~2.3$ & $44.2 \pm 0.9$ \\
Mrk\,507   & $ ~~~~5     $ & $0.30 \pm 0.03$ & $ 9.9 \pm~~ 5.8$ & $46.8 \pm 2.6$ \\
3C\,382	  & $ ~~15      $ & $0.24 \pm 0.03$ & $ 90.0 \pm~~3.3$ & $41.6 \pm 2.9$ \\
3C\,390.3  & $ ~~~~8     $ & $0.12 \pm 0.02$ & $112.7 \pm~~9.9$ & $29.0 \pm 2.6$ \\
NGC\,6814  & $ ~~60      $ & $0.08 \pm 0.01$ & $ 92.2 \pm 17.8$ & $23.6 \pm 1.6$ \\
Mrk\,509   & $ ~~~~8     $ & $0.16 \pm 0.02$ & $ 72.4 \pm~~3.6$ & $33.6 \pm 2.3$ \\
Mrk\,1513  & $ ~~13      $ & $0.54 \pm 0.01$ & $ 57.9 \pm~~1.2$ & $65.0 \pm 0.7$ \\
Mrk\,304   & $ ~~~~6     $ & $0.03 \pm 0.01$ & $ 54.5 \pm  26.5$ & $14.4 \pm 2.6$ \\
Ark\,564   & $ ~~16      $ & $0.22 \pm 0.01$ & $112.2 \pm~~1.9$ & $39.7 \pm 1.0$ \\
NGC\,7469  & $ ~~50      $ & $0.26 \pm 0.01$ & $124.4 \pm~~1.5$ & $43.4 \pm 0.9$ \\
Mrk\,315   & $ ~~13      $ & $0.14 \pm 0.01$ & $ 33.2 \pm~~5.4$ & $31.4 \pm 1.2$ \\
NGC\,7603  & $ ~~18      $ & $0.32 \pm 0.01$ & $166.8 \pm~~2.9$ & $48.4 \pm 0.8$ \\
Mrk\,541   & $ ~~20      $ & $0.36 \pm 0.01$ & $172.3 \pm~~1.3$ & $51.6 \pm 0.8$ \\

\hline
\end{tabular}
\end{center}
$^{\rm a}$ Start SMA, defining the region, over which the global values of the ellipticity and PA were estimated. The region generally extends to the profile end with two exceptions, for which the end SMA is given in parentheses.\\
$^{\rm b}$ The global parameters were taken from Simkin (\cite{S_75}).
\end{table}

\begin{table*}[t]
\caption{Total magnitudes.}
\label{t_totmag}
\begin{center}
\begin{tabular}{@{}llrrrr@{}}
\hline
\noalign{\smallskip}
~Galaxy    & ~Civil Date   & $B_{\rm tot}$~~~~~~ & $V_{\rm tot}$~~~~~~~~ & $R_{\rm C,tot}$~~~~~~\, & $I_{\rm C,tot}$~~~~~~~~ \\
          & (yyyy\,mm\,dd) & (mag)~~~~~\, & (mag)~~~~~\, & (mag)~~~~~\,  & (mag)~~~~~\, \\
\noalign{\smallskip}
\hline
\noalign{\smallskip}

 Mrk\,335 &  1998 08 22 &  $14.19 \pm 0.03$ &  $13.93 \pm 0.03$ &  $13.47 \pm 0.03$ &  $13.08 \pm 0.05$ \\	     
          &  2007 08 20 &  $  \ldots~~~~~~$ &  $14.02 \pm 0.02$ &  $13.59 \pm 0.02$ &  $13.30 \pm 0.04$ \\	     
III\,Zw\,2&  1997 09 09 &  $15.85 \pm 0.05$ &  $14.86 \pm 0.04$ &  $14.39 \pm 0.02$ &  $13.79 \pm 0.05$ \\
 Mrk\,348 &  1997 09 07 &  $13.79 \pm 0.05$ &  $12.90 \pm 0.04$ &  $12.38 \pm 0.04$ &  $11.86 \pm 0.04$ \\
 I\,Zw\,1 &  1998 08 20 &  $14.48 \pm 0.05$ &  $14.06 \pm 0.05$ &  $13.64 \pm 0.05$ &  $13.15 \pm 0.06$ \\
 Mrk\,352 &  2007 08 21 &  $  \ldots~~~~~~$ &  $  \ldots~~~~~~$ &  $13.82 \pm 0.01$ &  $  \ldots~~~~~~$ \\
          &  2008 01 30 &  $15.38 \pm 0.01$ &  $14.57 \pm 0.02$ &  $13.99 \pm 0.02$ &  $13.42 \pm 0.03$ \\
 Mrk\,573 &  1997 09 07 &  $13.97 \pm 0.06$ &  $13.17 \pm 0.04$ &  $12.67 \pm 0.04$ &  $12.07 \pm 0.05$ \\
 Mrk\,590 &  1997 09 06 &  $13.23 \pm 0.03$ &  $12.52 \pm 0.03$ &  $11.98 \pm 0.03$ &  $11.28 \pm 0.04$ \\
Mrk\,1040 &  1997 09 06 &  $13.65 \pm 0.03$ &  $12.80 \pm 0.03$ &  $12.19 \pm 0.03$ &  $11.73 \pm 0.04$ \\
 Mrk\,595 &  1997 09 09 &  $15.12 \pm 0.05$ &  $14.29 \pm 0.03$ &  $13.61 \pm 0.03$ &  $12.92 \pm 0.05$ \\
  3C\,120 &  1997 09 09 &  $14.55 \pm 0.04$ &  $13.86 \pm 0.03$ &  $13.19 \pm 0.03$ &  $12.51 \pm 0.05$ \\
          &  2008 02 01 &  $14.34 \pm 0.04$ &  $13.75 \pm 0.03$ &  $13.24 \pm 0.03$ &  $12.69 \pm 0.02$ \\
 Ark\,120 &  1994 09 29 &  $13.97 \pm 0.07$ &  $  \ldots~~~~~~$ &  $12.89 \pm 0.03$ &  $  \ldots~~~~~~$ \\
          &  1991 12 08 &  $  \ldots~~~~~~$ &  $13.45 \pm 0.05$ &  $  \ldots~~~~~~$ &  $  \ldots~~~~~~$ \\	 
 Mrk\,376 &  2008 02 03 &  $15.21 \pm 0.05$ &  $14.62 \pm 0.07$ &  $13.86 \pm 0.03$ &  $13.33 \pm 0.04$ \\
  Mrk\,79 &  1999 02 16 &  $13.72 \pm 0.03$ &  $13.08 \pm 0.03$ &  $12.48 \pm 0.03$ &  $11.91 \pm 0.05$ \\	     
          &  2008 02 01 &  $13.85 \pm 0.03$ &  $13.14 \pm 0.03$ &  $12.65 \pm 0.04$ &  $12.12 \pm 0.04$ \\
 Mrk\,382 &  1998 02 27 &  $15.30 \pm 0.03$ &  $14.63 \pm 0.03$ &  $14.09 \pm 0.04$ &  $13.56 \pm 0.05$ \\
          &  2008 02 02 &  $15.21 \pm 0.03$ &  $14.63 \pm 0.03$ &  $14.11 \pm 0.03$ &  $13.58 \pm 0.03$ \\
NGC\,3227 &  1999 04 17 &  $11.62 \pm 0.07$ &  $10.84 \pm 0.05$ &  $10.23 \pm 0.06$ &  $ 9.43 \pm 0.06$ \\
NGC\,3516 &  2008 01 08 &  $  \ldots~~~~~~$ &  $11.65 \pm 0.01$ &  $11.13 \pm 0.01$ &  $10.55 \pm 0.01$ \\
NGC\,4051 &  1995 05 06 &  $11.09 \pm 0.01$ &  $  \ldots~~~~~~$ &  $  \ldots~~~~~~$ &  $  \ldots~~~~~~$ \\ 	 
          &  2000 03 30 &  $  \ldots~~~~~~$ &  $  \ldots~~~~~~$ &  $ 9.78 \pm 0.02$ &  $  \ldots~~~~~~$ \\ 	 
          &  2001 04 09 &  $  \ldots~~~~~~$ &  $  \ldots~~~~~~$ &  $  \ldots~~~~~~$ &  $ 9.18 \pm 0.02$ \\ 	 
NGC\,4151 &  1999 03 10 &  $11.13 \pm 0.03$ &  $10.52 \pm 0.04$ &  $10.00 \pm 0.04$ &  $ 9.33 \pm 0.05$ \\
          &  1999 04 19 &  $11.25 \pm 0.05$ &  $10.56 \pm 0.04$ &  $10.05 \pm 0.04$ &  $ 9.45 \pm 0.05$ \\
 Mrk\,766 &  1999 02 15 &  $13.83 \pm 0.07$ &  $13.10 \pm 0.03$ &  $12.58 \pm 0.04$ &  $11.97 \pm 0.06$ \\
 Mrk\,771 &  1990 06 23 &  $  \ldots~~~~~~$ &  $14.37 \pm 0.03$ &  $  \ldots~~~~~~$ &  $13.11 \pm 0.01$ \\
NGC\,4593 &  2008 01 08 &  $  \ldots~~~~~~$ &  $11.11 \pm 0.01$ &  $10.59 \pm 0.01$ &  $10.00 \pm 0.01$ \\
 Mrk\,279 &  2008 02 02 &  $14.43 \pm 0.04$ &  $13.76 \pm 0.03$ &  $13.23 \pm 0.04$ &  $12.79 \pm 0.06$ \\
NGC\,5506 &  1999 04 17 &  $12.59 \pm 0.07$ &  $11.91 \pm 0.05$ &  $11.16 \pm 0.06$ &  $10.34 \pm 0.06$ \\
NGC\,5548 &  1999 04 19 &  $12.76 \pm 0.05$ &  $12.27 \pm 0.04$ &  $11.79 \pm 0.05$ &  $11.18 \pm 0.06$ \\
 Ark\,479 &  2007 07 19 &  $  \ldots~~~~~~$ &  $14.62 \pm 0.06$ &  $14.05 \pm 0.07$ &  $13.48 \pm 0.08$ \\
 Mrk\,506 &  1997 06 01 &  $15.09 \pm 0.04$ &  $14.29 \pm 0.04$ &  $13.76 \pm 0.05$ &  $13.01 \pm 0.05$ \\
          &  1998 07 18 &  $15.01 \pm 0.04$ &  $14.20 \pm 0.02$ &  $13.62 \pm 0.03$ &  $12.99 \pm 0.05$ \\
          &  2007 06 17 &  $15.38 \pm 0.03$ &  $14.33 \pm 0.02$ &  $13.82 \pm 0.03$ &  $  \ldots~~~~~~$ \\ 
 Mrk\,507 &  1998 07 20 &  $16.77 \pm 0.06$ &  $15.85 \pm 0.05$ &  $15.21 \pm 0.05$ &  $14.23 \pm 0.06$ \\
  3C\,382 &  1998 08 23 &  $14.18 \pm 0.04$ &  $13.85 \pm 0.04$ &  $13.31 \pm 0.04$ &  $12.78 \pm 0.05$ \\
3C\,390.3 &  1998 08 20 &  $15.93 \pm 0.06$ &  $15.05 \pm 0.04$ &  $14.40 \pm 0.04$ &  $13.93 \pm 0.05$ \\
NGC\,6814 &  1997 07 06 &  $12.16 \pm 0.02$ &  $11.30 \pm 0.02$ &  $10.37 \pm 0.02$ &  $ 9.62 \pm 0.03$ \\
          &  1997 07 10 &  $12.30 \pm 0.03$ &  $11.36 \pm 0.02$ &  $10.56 \pm 0.03$ &  $ 9.69 \pm 0.04$ \\
          &  1997 09 07 &  $12.14 \pm 0.05$ &  $11.10 \pm 0.03$ &  $10.36 \pm 0.03$ &  $ 9.49 \pm 0.04$ \\
          &  1998 07 18 &  $12.28 \pm 0.04$ &  $11.26 \pm 0.02$ &  $10.54 \pm 0.03$ &  $ 9.79 \pm 0.05$ \\
 Mrk\,509 &  1997 07 10 &  $14.09 \pm 0.03$ &  $13.59 \pm 0.02$ &  $13.08 \pm 0.03$ &  $12.65 \pm 0.04$ \\
          &  1997 09 08 &  $13.89 \pm 0.03$ &  $13.48 \pm 0.02$ &  $12.99 \pm 0.03$ &  $12.69 \pm 0.04$ \\
          &  1998 07 20 &  $13.40 \pm 0.05$ &  $13.16 \pm 0.04$ &  $12.70 \pm 0.04$ &  $12.30 \pm 0.05$ \\
Mrk\,1513 &  2007 08 20 &  $  \ldots~~~~~~$ &  $14.54 \pm 0.02$ &  $14.08 \pm 0.02$ &  $13.79 \pm 0.01$ \\
 Mrk\,304 &  1998 07 19 &  $14.61 \pm 0.04$ &  $14.31 \pm 0.03$ &  $13.82 \pm 0.03$ &  $13.50 \pm 0.04$ \\
 Ark\,564 &  1998 07 18 &  $14.22 \pm 0.04$ &  $13.71 \pm 0.02$ &  $13.27 \pm 0.03$ &  $12.97 \pm 0.04$ \\
          &  1998 08 20 &  $14.34 \pm 0.05$ &  $13.82 \pm 0.04$ &  $13.37 \pm 0.04$ &  $12.90 \pm 0.05$ \\
NGC\,7469 &  1997 09 06 &  $12.73 \pm 0.03$ &  $12.17 \pm 0.03$ &  $11.67 \pm 0.02$ &  $11.01 \pm 0.04$ \\
          &  1998 07 19 &  $12.70 \pm 0.04$ &  $12.18 \pm 0.03$ &  $11.77 \pm 0.03$ &  $11.11 \pm 0.04$ \\
          &  1998 08 23 &  $12.87 \pm 0.04$ &  $12.40 \pm 0.04$ &  $11.80 \pm 0.04$ &  $11.05 \pm 0.05$ \\
          &  2003 07 28 &  $13.33 \pm 0.03$ &  $12.42 \pm 0.03$ &  $11.75 \pm 0.02$ &  $11.45 \pm 0.03$ \\       
 Mrk\,315 &  2007 08 22 &  $  \ldots~~~~~~$ &  $  \ldots~~~~~~$ &  $13.89 \pm 0.03$ &  $  \ldots~~~~~~$ \\
NGC\,7603 &  2007 07 19 &  $  \ldots~~~~~~$ &  $12.47 \pm 0.07$ &  $  \ldots~~~~~~$ &  $11.51 \pm 0.08$ \\
 Mrk\,541 &  2007 07 19 &  $  \ldots~~~~~~$ &  $14.73 \pm 0.07$ &  $14.12 \pm 0.07$ &  $13.52 \pm 0.08$ \\

\hline
\end{tabular}
\end{center}
\end{table*}

\begin{figure*}[t]
\centering
 \begin{minipage}[t]{5cm}   
\includegraphics[width=5cm]{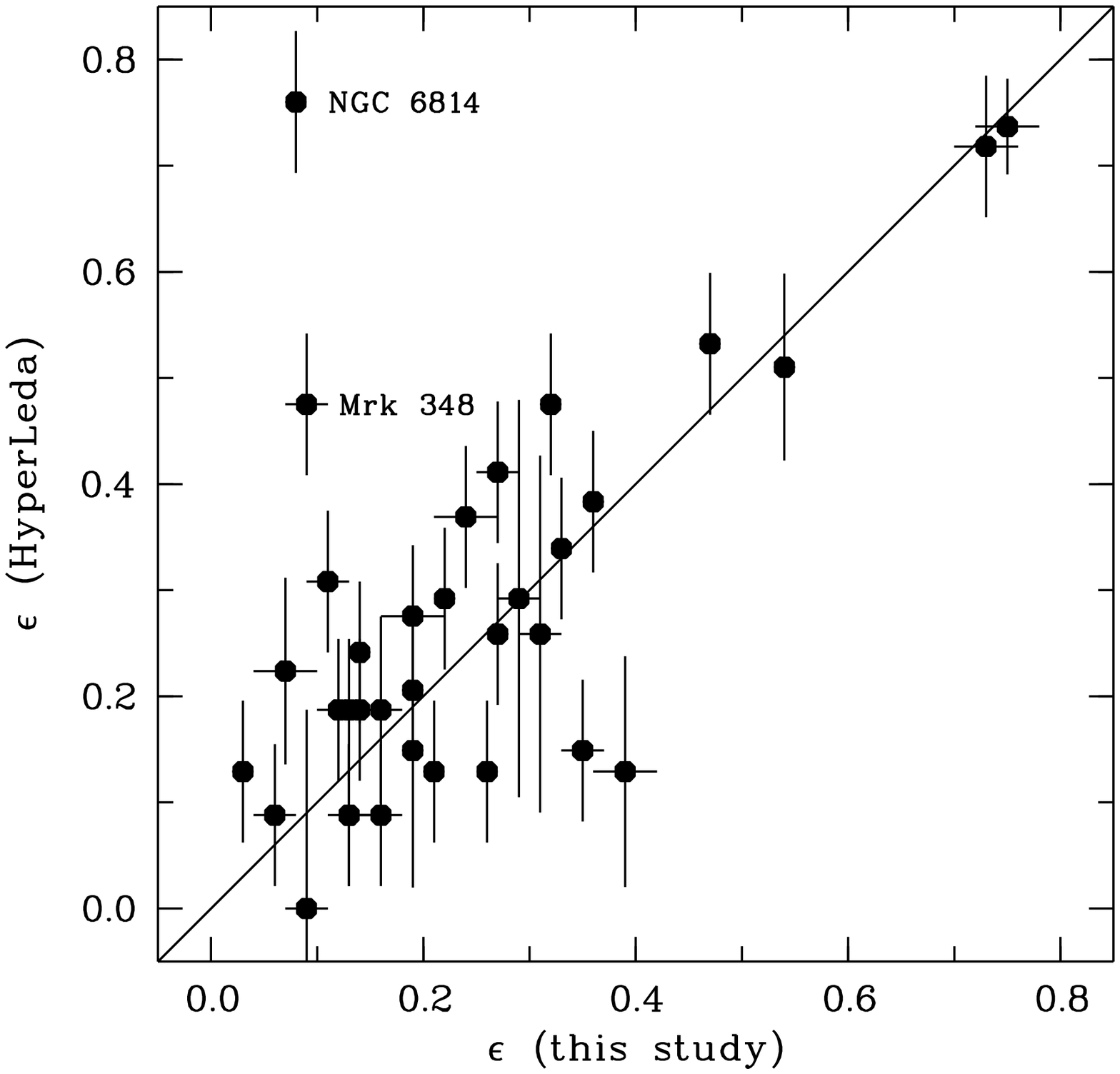}
 \caption{Comparison between the ellipticities listed in HyperLeda and ours. Named are the galaxies with $|\Delta\epsilon|\,$$>$$\,0.3$. The line of exact correspondence is plotted.}
 \label{E}
 \end{minipage}
 \hspace{0.5cm}
\begin{minipage}[t]{5cm}   
\includegraphics[width=5cm]{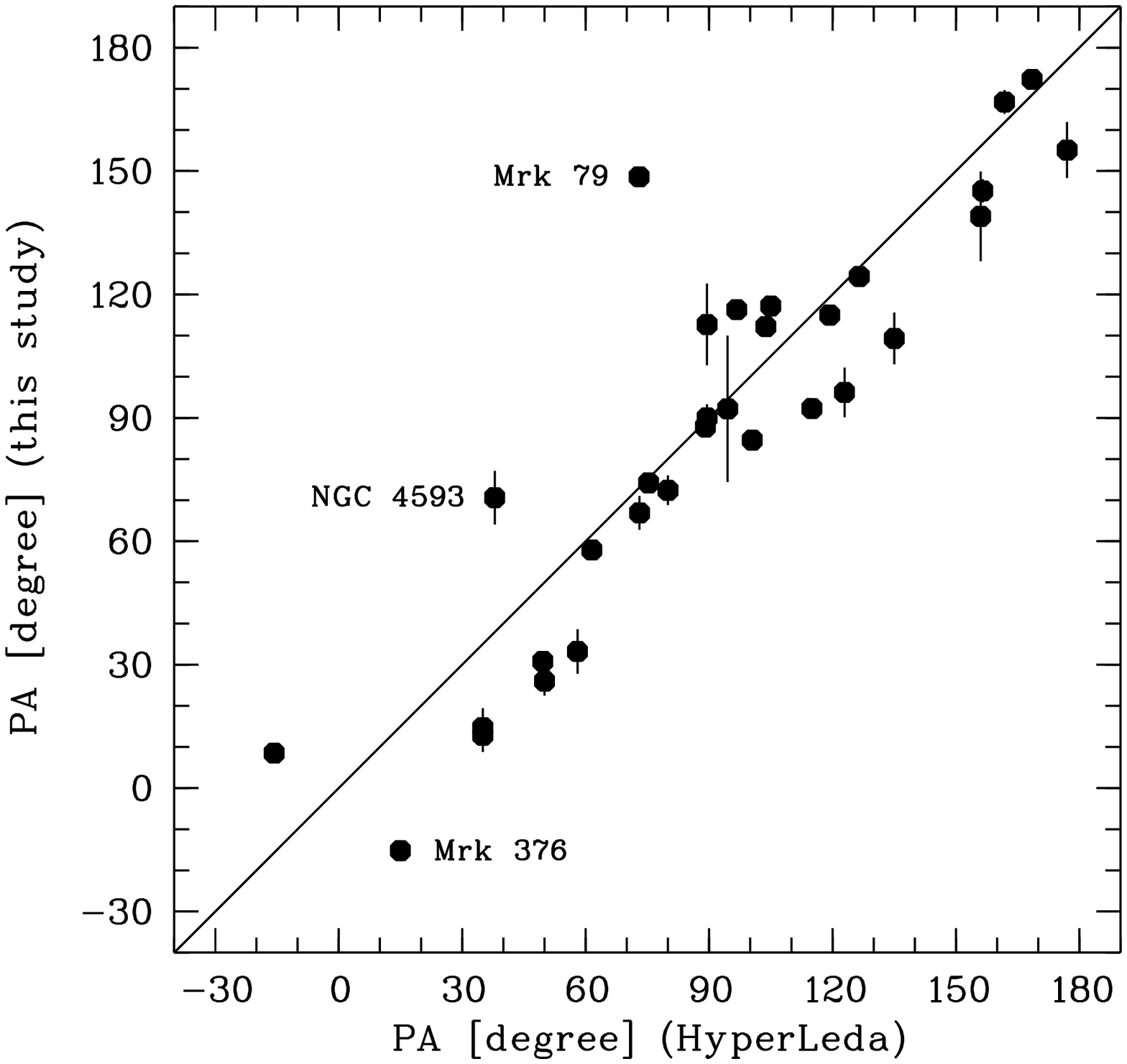}
\caption{Comparison between the PAs listed in HyperLeda and ours. Named are the galaxies with $|\Delta\rm PA|\,$$>$$\,30\degr$. The line of exact correspondence is plotted.}
\label{PA}
\end{minipage}
\end{figure*}

\begin{figure*}[t]
\centering
\begin{minipage}[t]{5cm}   
\includegraphics[width=5cm]{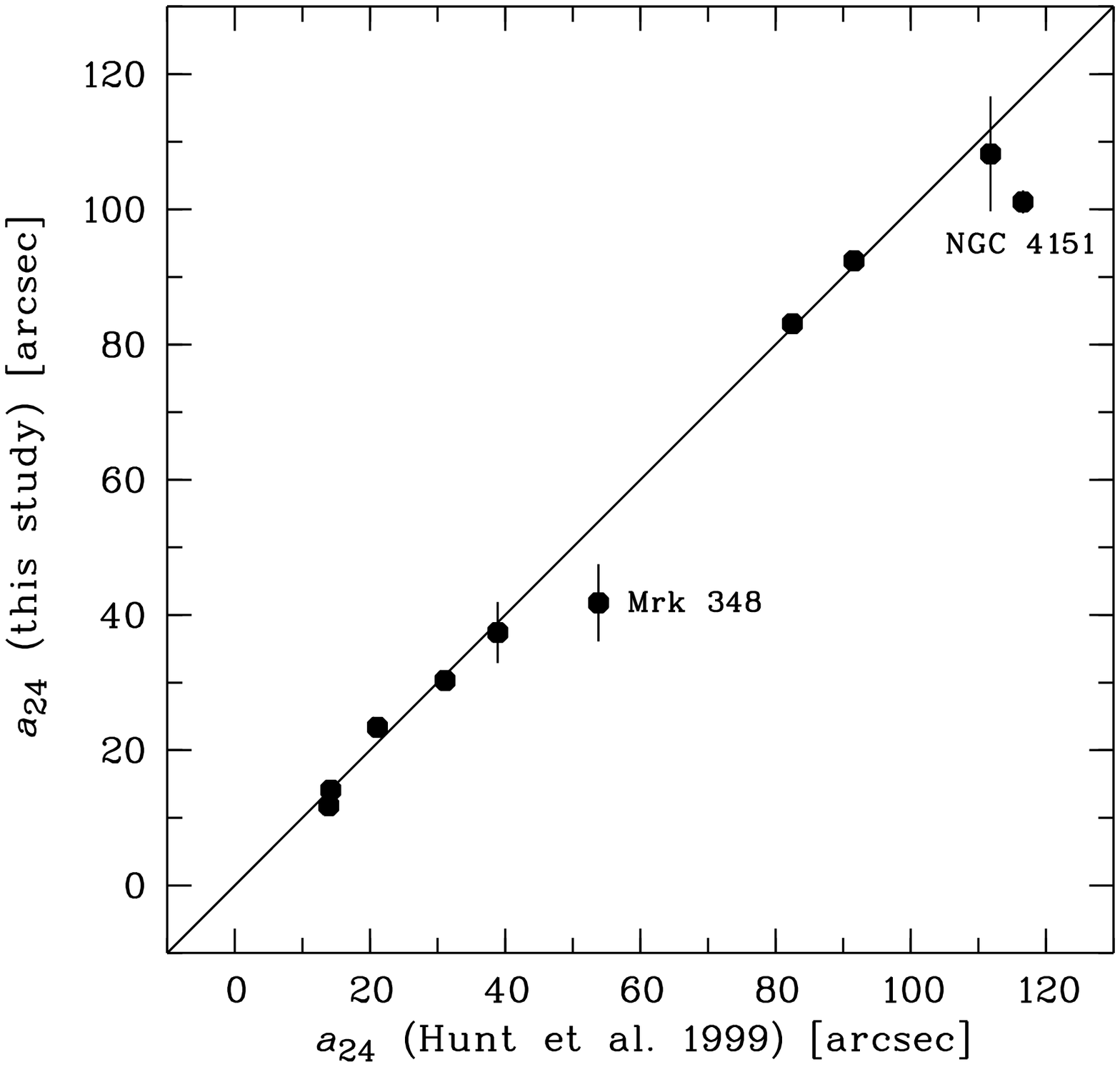}
\caption{
Comparison between $a_{24}$ estimated by us and those published in Hunt et~al. (\cite{HMR2_99}). Named are the outliers. The line of exact correspondence is plotted.}
\label{comp_isosiz}
\end{minipage}
\hspace{0.5cm}
\begin{minipage}[t]{5cm}   
\includegraphics[width=5cm]{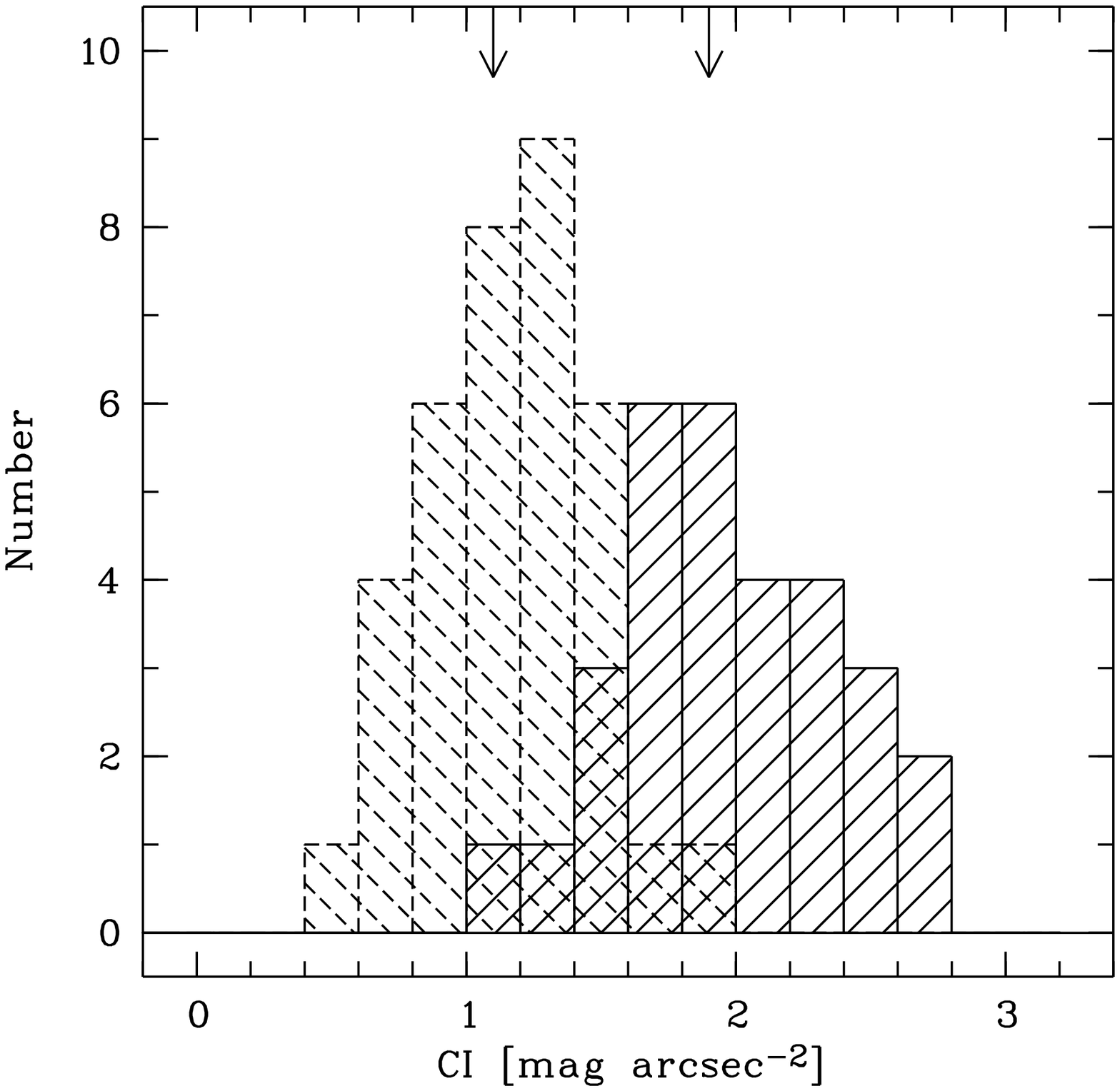}
\caption{Distribution of $(B$$-$$I_{\rm \scriptstyle C})_{24}^{(0)}$ (solid) and $(V$$-$$I_{\rm \scriptstyle C})_{24}^{(0)}\,$ (dashed), whose median values are denoted by the right and left arrow, respectively.}
\label{col_hist}
\end{minipage}
\hspace{0.5cm}
\begin{minipage}[t]{5cm}   
\includegraphics[width=5cm]{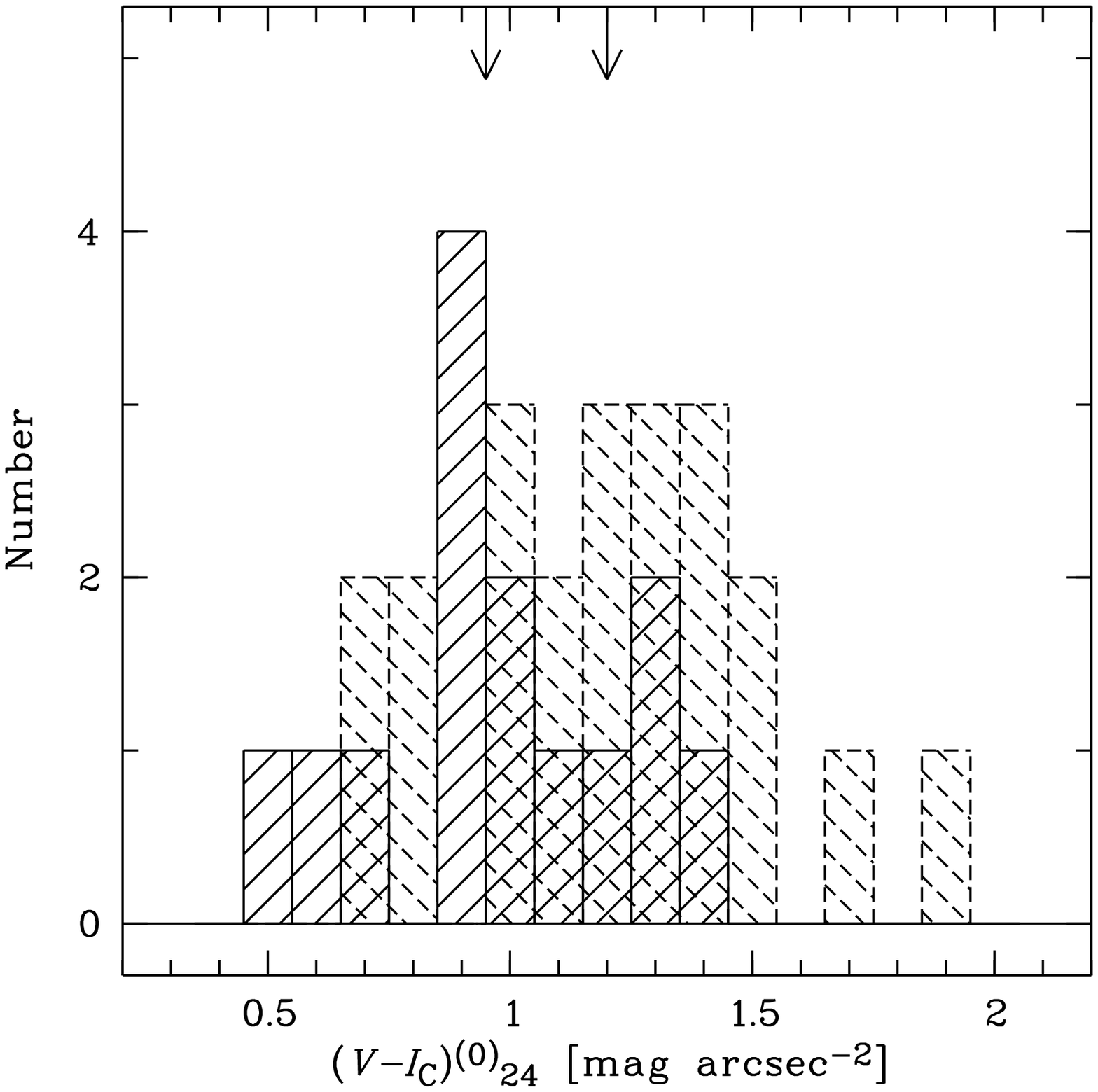}
\caption{Distribution of $(V$$-$$I_{\rm \scriptstyle C})_{24}^{(0)}$ for the galaxies with/without outer rings (solid/dashed)  with the median values denoted by the left/right arrow.}
\label{comp_isosize}
\end{minipage}
\end{figure*}

\section{Global parameters}
\label{glopar}

We present the global values of the ellipticity and PA of the sample galaxies in Table\,\ref{T_glopar}. They were obtained
computing the median over a predefined region of the corresponding profiles; the mean absolute deviation
about the median (MAD) was used as an error estimate.
The region of estimations is one and the same for all passbands and epochs for a given
galaxy and generally encompasses the disk-dominated parts (Table\,\ref{T_glopar}, see also Fig.\,A1 in Paper\,I).
All available passbands were considered in the median computation for a given epoch. Multi-epoch global parameters were weight-averaged; the galaxies of multi-epoch observations can be followed in Table\,3 of Paper\,I.

Having estimated the ellipticity, we obtained the galaxy inclination, $i$, using the expression
(Holmberg \cite{H_58}): 
\begin{displaymath}
\cos i=\sqrt{{(1\!-\!\epsilon)^{2}-q_{0}^{2} \over 1-q_{0}^{2}}\,}, 
\end{displaymath}
where $q_0\,$$=$$\,0.2$ is the intrinsic, edge-on disk axial ratio (Lambas, Maddox \& Loveday \cite{LML_92}). The so derived inclination is presented in Table\,\ref{T_glopar}.
The outermost, disk-dominated, parts of NGC\,4151 could not be reached in our images, so, we list the parameters estimated by Simkin (\cite{S_75}).

The typical error of the global PA is a few degrees, but it may get higher for nearly face-on galaxies ($\epsilon$$\,\la$\,0.1) as no favourable PA could be defined (e.g., Mrk\,590, see Paper\,I) and owing to the presence of spiral arms as they can lead to a continuous PA change (e.g., Mrk\,348, see Paper\,I).
We found good passband-to-passband and night-to-night correspondence of the global ellipticities and PAs as can be judged by the small values of the errors reported in Table\,\ref{T_glopar}. 

We compared the values of the global ellipticity and PA with those listed in HyperLeda\footnote{http://leda.univ-lyon1.fr} (\cite{PPP_03}) in Figs.\,\ref{E} and \ref{PA}; note that HyperLeda reports the parameters corresponding to the 25 $B$ mag arcsec$^{-2}$ isophote. The median values of the difference of the ellipticity and PA (ours minus theirs) are $-0.02$ (0.10) and  $-4\fdg3$ ($15\fdg8$), respectively, with MAD in parentheses.
Two galaxies have ellipticity difference $|\Delta\epsilon|\,$$>$$\,0.3$~-- NGC\,6814 and Mrk\,348. The former, practically a face-on galaxy, is reported to have $\epsilon$$\,=\,$0.76. The latter is also an almost face-on galaxy with a stretched spiral structure beyond $a$$\,\approx$$\,40^{\prime\prime}$, where the 25 $B$ mag arcsec$^{-2}$ isophote gets (see Fig.\,A.1 of Paper\,I).
Similar is the case of Mrk\,79, at the top of the list of galaxies with PA difference $|\Delta\rm PA|\,$$>$$\,30\degr$ ~-- the 25 $B$ mag arcsec$^{-2}$ isophote falls in the region of a continuous PA change owing to the spiral arms (see Fig.\,A.1 of Paper\,I).
Thus, the relatively large scatter in Figs.\,\ref{E} and \,\ref{PA} is associated with the different way of estimation of these parameters.
Furthermore, the ellipticity (together with inclination) and PA, estimated on the base of isophotal criteria, are not fully representative for the disk-dominated galaxy regions.

The global ellipticities of the matched inactive galaxies are presented in Appendix\,\ref{AppendixB}.

We built the growth curve for each galaxy by integrating the intensity in elliptical apertures of fixed centre, ellipticity, and PA, using the global ellipticity and PA. The intensity, at which the growth curve becomes asymptotically flat, specifies the total apparent magnitude, presented in Table\,\ref{t_totmag}. The total magnitude error includes photon noise, sky background error, and transformation coefficient errors.
No magnitude averaging was performed, so that eventual changes related to nuclear variability could remain.

\begin{figure*}[t]
\centering
\begin{minipage}[t]{5cm}   
\includegraphics[width=5cm]{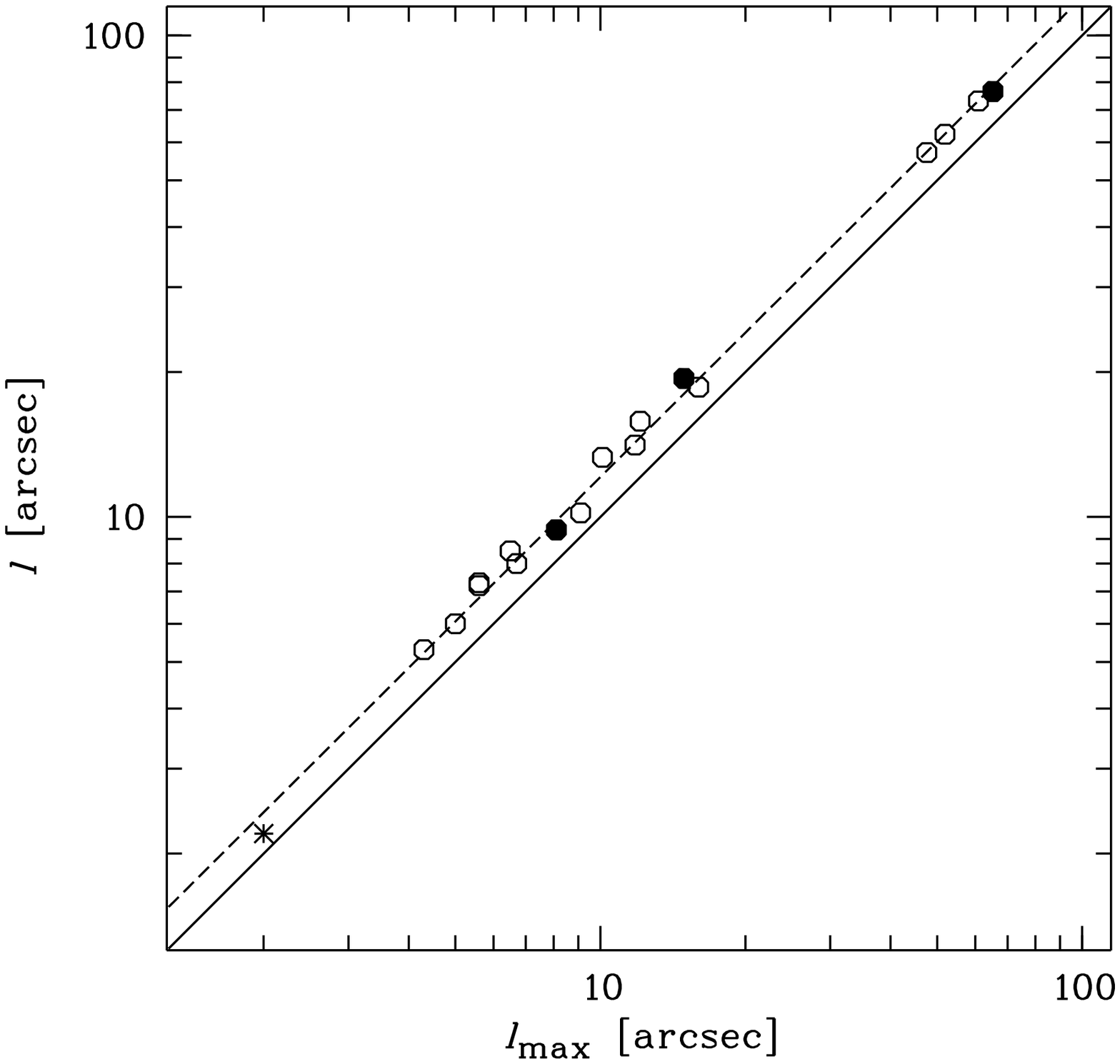}
\caption{
Bar length vs. bar SMA at ellipticity maximum of strong (filled circles) and weak (open circles) bars; the nuclear bar of Mrk\,352 is denoted by an asterisk. 
Overplotted are the lines of the median $\ell/\ell_{\rm max}$ ratio (dashed) and of exact correspondence (solid).}
\label{B_LmaxLa_lgr}
\end{minipage}
\hspace{0.5cm}
\begin{minipage}[t]{5cm}   
\includegraphics[width=5cm]{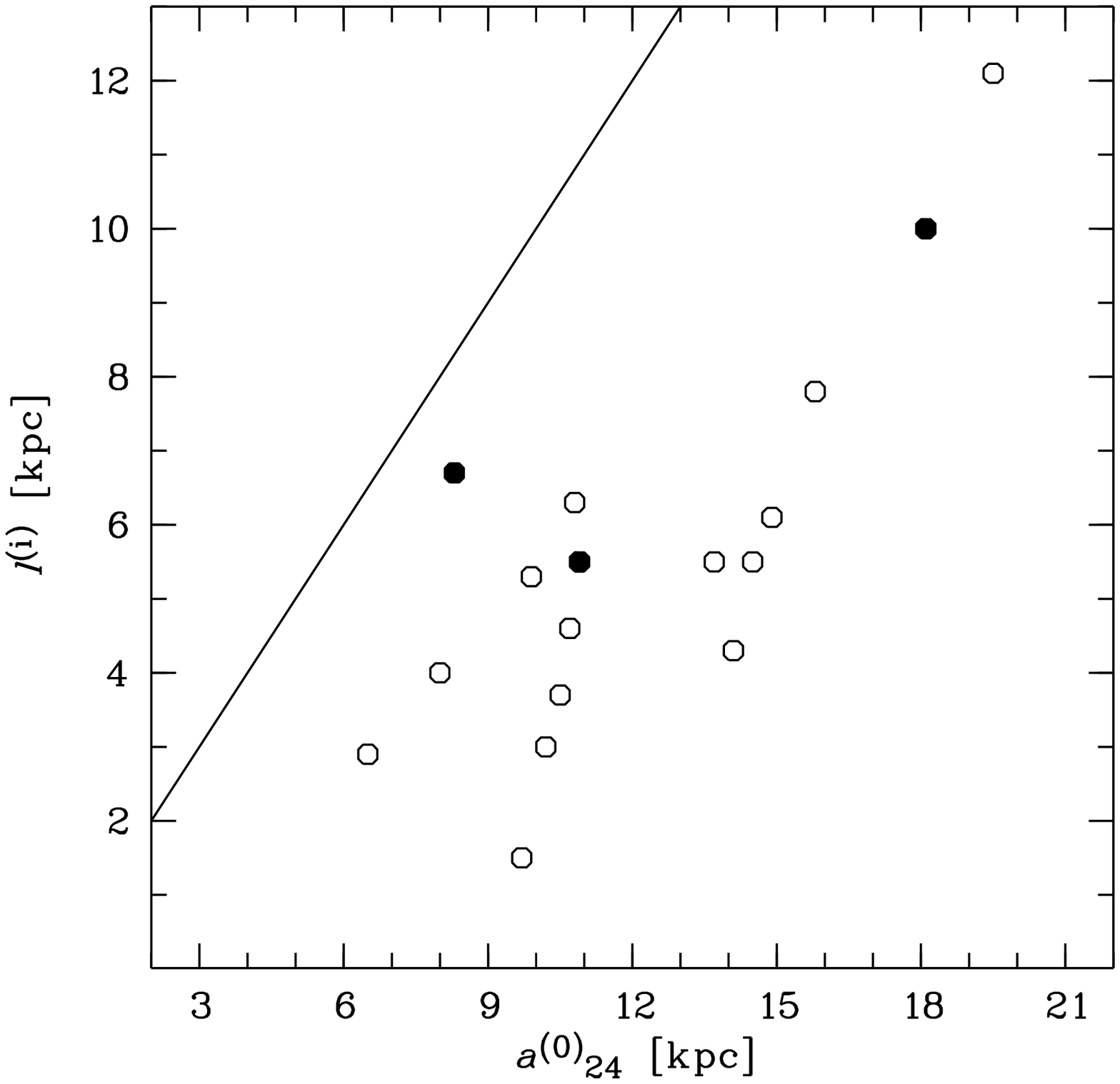}
\caption{
Deprojected bar length vs. corrected isophotal SMA at 24 $V$ mag arcsec$^{-2}$; the symbols are as specified in Fig.\,\ref{B_LmaxLa_lgr}. 
Overplotted is the line of exact correspondence.}
\label{B_AvLi}
\end{minipage}
\hspace{0.5cm}
\begin{minipage}[t]{5cm}   
\includegraphics[width=5cm]{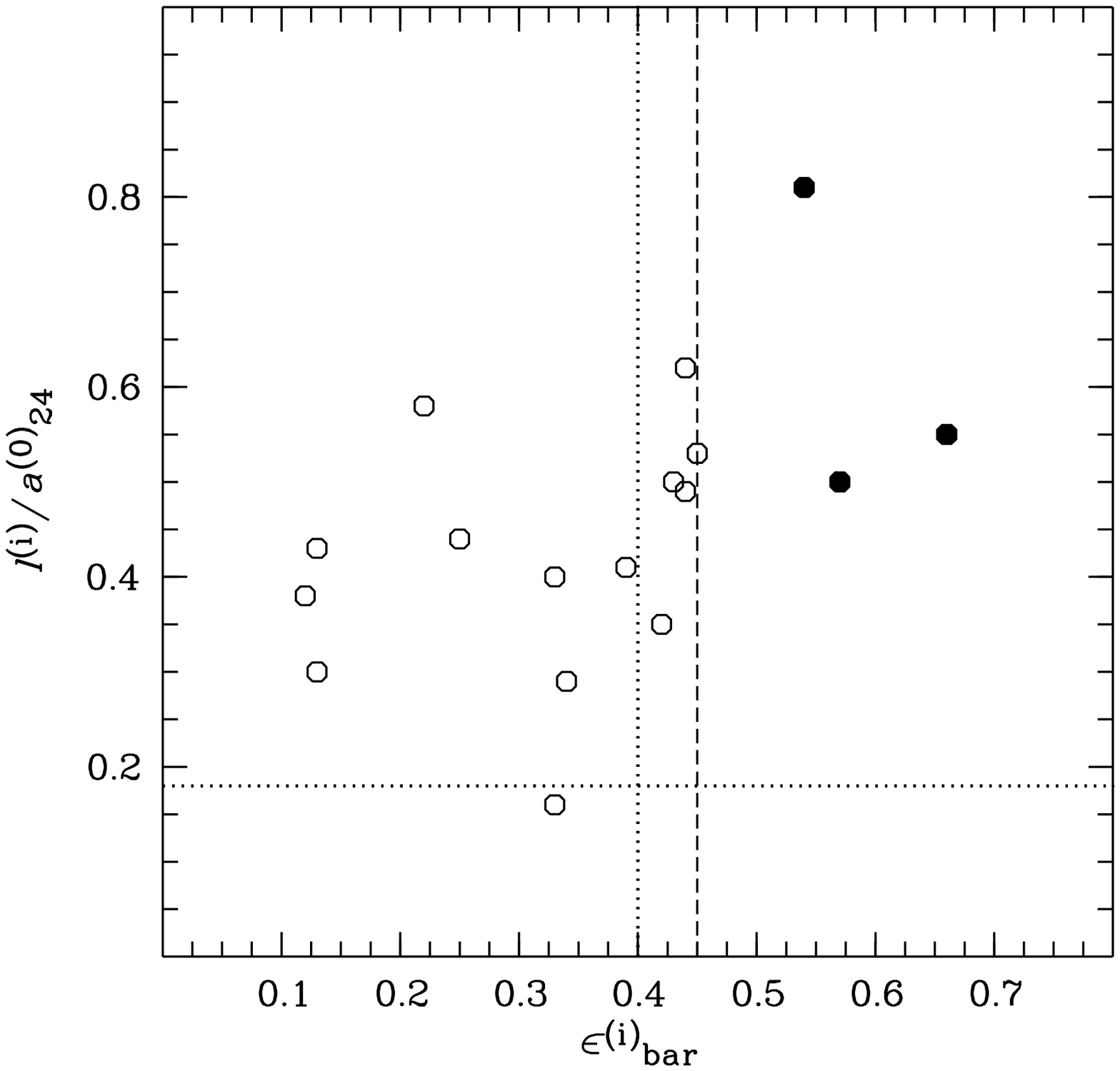}
\caption{
Deprojected relative bar length vs. deprojected bar ellipticity; the symbols are as specified in Fig.\,\ref{B_LmaxLa_lgr}. 
Overplotted are empirical limits of strong and long bars (see text).}
\label{B_EiLni}
\end{minipage}
\end{figure*}

\section{Isophotal parameters} 
\label{isopar}

In Table\,\ref{t_isopar} we present a set of 24 $V$ mag arcsec$^{-2}$ isophote parameters: the SMA $a_{24}$ together with the CIs $(B\,$--$\,I_{\rm \scriptstyle C})_{24}$ and $(V\,$--$\,I_{\rm \scriptstyle C})_{24}$ estimated on both the apparent, $\mu(a)$, and corrected, $\mu^{(0)}(a)$, SB profiles; the superscript ``0'' denotes the corrected quantities. Correction was performed as follows:
\begin{displaymath}
\mu^{(0)}(a)=\mu(a)-\Sigma,
\end{displaymath}
where 
\begin{displaymath}
\Sigma=A+2.5\,C\log(\cos i)+10\log(1+z)+K+E
\end{displaymath}
accounts for the following factors:

\begin{itemize}

\item[--] {\em Galactic absorption, A.} Its values calculated following Schlegel, Finkbeiner \& Davis (\cite{SFD_98}) were taken from NED;

\item[--] {\em Correction for inclination and internal absorption.} 
Factor $C$ accounts for the disk transparency owing to dust absorption. It depends on the radial distance and is assumed to vary from about 0 in the disk centre to about 1 in the galaxy outskirts (e.g., Giovanelli et~al. \cite{GHS_94}). For the sake of estimating the 24 $V$ mag arcsec$^{-2}$ parameters, we adopted $C\,$$=$$\,0.9$ (Nedyalkov \cite{N_98});

\item[--] {\em Cosmological dimming};

\item[--] {\em $K$- and $E$-correction.} We used the following fits to the data points with $z\,$$\le$\,0.1 and Hubble type Sa of Poggianti (\cite{P_97}):
\begin{eqnarray}
K_{B}&=&4.290\,z, \nonumber \\
K_{V}&=&1.370\,z, \nonumber \\
K_{I}&=&-0.014\,z+1.786\,z^{2}, \nonumber
\end{eqnarray} 
and 
\begin{eqnarray}
E_{B}&=&-2.470\,z, \nonumber \\
E_{V}&=&-1.945\,z, \nonumber \\
E_{I}&=&-1.615\,z. \nonumber 
\end{eqnarray} 
Poggianti (\cite{P_97}) did not tabulate $K$- and $E$-corrections for Sa in the Cousins $I$, so, we used the corrections for the Johnson $I$.

\end{itemize}

When the 24 $V$ mag arcsec$^{-2}$ isophote could not be reached, extrapolation, instead of interpolation, was perfor\-med on the base of a linear fit to the disk-dominated galaxy regions, using the following formula (Han \cite{H_92}):
\begin{displaymath}
a_{24}^{\rm (x)}=(24-\mu_{0})\,{r_{0}\over1.0857},
\end{displaymath}
where $a_{24}^{\rm (x)}$ is the extrapolated $a_{24}$ and $\mu_{0}$ and $r_{0}$ the central SB and the scale length of the fitted galaxy disk, respectively. In case the CI profiles do not reach $a_{24}$, linear fits to the corresponding SB profiles were used to extrapolate them. The multi-epoch isophotal parameters were weight-averaged. 

Our values of $a_{24}$ are in good agreement with those of Hunt et~al. (\cite{HMR2_99}) for the galaxies in common (Fig.\,\ref{comp_isosiz}).
The median value of the difference (ours minus theirs) is  $-1\farcs2$ with MAD of  $3\farcs8$.
The outliers are Mrk\,348 and NGC\,4151. Possible reasons for this are eventual differences in the sky background estimation and that, at variance with us, Hunt et~al. (\cite{HMR2_99}) used fixed values of the ellipticity and PA in the ellipse fitting in the region of the 24 $V$ mag arcsec$^{-2}$ isophote for most galaxies, including the above cited. Note also that in this region the SB profile of Mrk\,348 has  an almost flat appearance (see Fig.\,A1 in Paper\,I), which results in a relatively large error in the estimation of $a_{24}$ and that Hunt et~al. (\cite{HMR2_99}) applied extrapolation to estimate $a_{24}$ of NGC\,4151.

The median corrected isophotal SMA is 13.9\,kpc with MAD of 3.7\,kpc.
The distribution of the corrected CIs is shown in Fig.\,\ref{col_hist}.
We acquired the following median CIs:
\begin{eqnarray}
(B-I_{\rm \scriptstyle C})_{24}&=&2.2\,(0.4)\,\rm mag\,arcsec^{-2}, \nonumber \\
(V-I_{\rm \scriptstyle C})_{24}&=&1.3\,(0.2)\,\rm mag\,arcsec^{-2}, \nonumber \\
(B-I_{\rm \scriptstyle C})_{24}^{(0)}&=&1.9\,(0.3)\,\rm mag\,arcsec^{-2}, \nonumber \\
(V-I_{\rm \scriptstyle C})_{24}^{(0)}&=&1.1\,(0.3)\,\rm mag\,arcsec^{-2}, \nonumber 
\end{eqnarray}
which are representative for the disk-dominated galaxy regions; the values in parentheses are the MADs.

We compared the distribution of the corrected CIs in the subsamples that have bars, outer/inner rings, asymmetries, companions and in the corresponding subsamples without these features. 
The subsample with outer rings shows bluer CIs than the complementary subsample 
at more than the 95\% significance level. In particular, the median $(V-I_{\rm \scriptstyle C})_{24}^{(0)}$ of the subsample with/without outer rings is 0.95/1.20 mag arcsec$^{-2}$ (Fig.\,\ref{comp_isosize}). This is expected, since outer rings, typically blue, are  generally situated close to the 24 $V$ mag arcsec$^{-2}$ isophote.
There is no clear correlation between the CIs and the presence of the other features.

\begin{table*}[t]
\caption{Apparent and corrected SMA and CIs at 24 $V$ mag arcsec$^{-2}$.}
\label{t_isopar}
\begin{center}
\begin{tabular}{@{}lllllll@{}}
\hline
\noalign{\smallskip}
~Galaxy & ~~~~~~~~$a_{24}$ & ~~~~$(B-I_{\rm \scriptstyle C})_{24}$ & ~~~~$(V-I_{\rm \scriptstyle C})_{24}$ & ~~~~~~~$a_{24}^{(0)}$ & ~~~~$(B-I_{\rm \scriptstyle C})_{24}^{(0)}$ & ~~~~$(V-I_{\rm \scriptstyle C})_{24}^{(0)}$ \\
\noalign{\smallskip}
 & ~~~~(arcsec) & ($\rm mag\,arcsec^{-2}$) & ($\rm mag\,arcsec^{-2}$) & ~~~~~(kpc) & ($\rm mag\,arcsec^{-2}$) & ($\rm mag\,arcsec^{-2}$) \\
\noalign{\smallskip}
\hline
\noalign{\smallskip}

Mrk\,335            & $~~11.8 \pm 0.7            $ & ~~~~~$2.3 \pm 0.3          $ & ~~~~~$1.5 \pm 0.3          $ & $~~5.9 \pm 0.4            $ & ~~~~~$2.1 \pm 0.3          $ & ~~~~~$1.4 \pm 0.3          $ \\
III\,Zw\,2          & $~~10.8 \pm 0.5            $ & ~~~~~$3.2 \pm 0.2          $ & ~~~~~$1.6 \pm 0.2          $ & $ 20.5 \pm 1.8            $ & ~~~~~$2.7 \pm 0.2          $ & ~~~~~$1.4 \pm 0.2          $ \\ 
Mrk\,348            & $~~41.8 \pm 5.7            $ & ~~~~~$1.8 \pm 0.3          $ & ~~~~~$0.9 \pm 0.2          $ & $ 15.5 \pm 4.3            $ & ~~~~~$1.6 \pm 0.3          $ & ~~~~~$0.8 \pm 0.2          $ \\ 
I\,Zw\,1            & $~~14.1 \pm 0.7            $ & ~~~~~$2.1 \pm 0.2          $ & ~~~~~$1.3 \pm 0.2          $ & $ 17.0 \pm 0.8            $ & ~~~~~$1.8 \pm 0.2          $ & ~~~~~$1.1 \pm 0.2          $ \\ 
Mrk\,352            & $~~15.6 \pm 0.5            $ & ~~~~~$2.2 \pm 0.3          $ & ~~~~~$1.3 \pm 0.2          $ & $~~4.2 \pm 0.1            $ & ~~~~~$2.0 \pm 0.3          $ & ~~~~~$1.2 \pm 0.2          $ \\ 
Mrk\,573            & $~~33.3 \pm 1.7            $ & ~~~~~$1.9 \pm 0.4          $ & ~~~~~$1.0 \pm 0.2          $ & $ 10.5 \pm 0.6            $ & ~~~~~$1.7 \pm 0.4          $ & ~~~~~$0.9 \pm 0.2          $ \\ 
Mrk\,590            & $~~40.2 \pm 1.8            $ & ~~~~~$2.9 \pm 0.3          $ & ~~~~~$1.3 \pm 0.2          $ & $ 20.8 \pm 0.9            $ & ~~~~~$2.7 \pm 0.3          $ & ~~~~~$1.3 \pm 0.2          $ \\ 
Mrk\,1040           & $~~83.1 \pm 0.4^{\,\rm x}  $ & ~~~~~$1.6 \pm 0.5          $ & ~~~~~$1.0 \pm 0.9^{\,\rm x}$ & $ 16.7 \pm 3.2            $ & ~~~~~$1.4 \pm 0.5          $ & ~~~~~$0.8 \pm 0.9^{\,\rm x}$ \\ 
Mrk\,595            & $~~20.5 \pm 0.8            $ & ~~~~~$2.7 \pm 0.2          $ & ~~~~~$1.6 \pm 0.2          $ & $ 10.7 \pm 0.4            $ & ~~~~~$2.3 \pm 0.2          $ & ~~~~~$1.4 \pm 0.2          $ \\ 
3C\,120             & $~~23.4 \pm 1.5            $ & ~~~~~$2.5 \pm 0.3^{\,\rm x}$ & ~~~~~$1.4 \pm 0.2          $ & $ 18.2 \pm 2.3            $ & ~~~~~$1.7 \pm 0.3^{\,\rm x}$ & ~~~~~$1.0 \pm 0.2          $ \\ 
Ark\,120            & $~~18.4 \pm 0.2^{\,\rm x}  $ & ~~~~~~~~~~$    \ldots      $ & ~~~~~~~~~~$    \ldots      $ & $ 13.1 \pm 0.1^{\,\rm x}  $ & ~~~~~~~~~~$    \ldots      $ & ~~~~~~~~~~$    \ldots      $ \\ 
Mrk\,376            & $~~13.6 \pm 1.2            $ & ~~~~~$2.3 \pm 0.8^{\,\rm x}$ & ~~~~~$1.4 \pm 0.4          $ & $ 14.9 \pm 1.5            $ & ~~~~~$1.9 \pm 0.8^{\,\rm x}$ & ~~~~~$1.2 \pm 0.4          $ \\ 
Mrk\,79             & $~~37.4 \pm 4.5            $ & ~~~~~$1.8 \pm 0.3          $ & ~~~~~$0.9 \pm 0.3          $ & $ 18.1 \pm 1.7            $ & ~~~~~$1.6 \pm 0.3          $ & ~~~~~$0.7 \pm 0.3          $ \\ 
Mrk\,382            & $~~20.5 \pm 0.7            $ & ~~~~~$1.8 \pm 0.3          $ & ~~~~~$0.8 \pm 0.2          $ & $ 13.7 \pm 0.5            $ & ~~~~~$1.5 \pm 0.3          $ & ~~~~~$0.7 \pm 0.2          $ \\ 
NGC\,3227           & $ 127.3 \pm 1.0^{\,\rm x}  $ & ~~~~~$2.4 \pm 0.4^{\,\rm x}$ & ~~~~~$1.9 \pm 0.3^{\,\rm x}$ & $ 10.8 \pm 0.1^{\,\rm x}  $ & ~~~~~$2.4 \pm 0.4^{\,\rm x}$ & ~~~~~$1.9 \pm 0.3^{\,\rm x}$ \\ 
NGC\,3516           & $~~57.7 \pm 4.5            $ & ~~~~~~~~~~$    \ldots      $ & ~~~~~$1.0 \pm 0.4          $ & $ 10.2 \pm 0.8            $ & ~~~~~~~~~~$    \ldots      $ & ~~~~~$0.9 \pm 0.4          $ \\ 
NGC\,4051           & $ 181.9 \pm 5.5^{\,\rm a,x}$ & ~~~~~$2.4 \pm 0.9^{\,\rm x}$ & ~~~~~~~~~~$    \ldots      $ & $~~9.9 \pm 0.3^{\,\rm a,x}$ & ~~~~~$2.3 \pm 0.9^{\,\rm  x}$ & ~~~~~~~~~~$    \ldots      $ \\ 
NGC\,4151           & $ 101.1 \pm 1.7            $ & ~~~~~$2.0 \pm 0.5          $ & ~~~~~$1.3 \pm 0.5          $ & $~~8.3 \pm 0.3            $ & ~~~~~$1.9 \pm 0.5          $ & ~~~~~$1.3 \pm 0.5          $ \\ 
Mrk\,766            & $~~30.3 \pm 1.0            $ & ~~~~~$2.0 \pm 0.2          $ & ~~~~~$1.3 \pm 0.2          $ & $~~8.0 \pm 0.2            $ & ~~~~~$1.9 \pm 0.2          $ & ~~~~~$1.3 \pm 0.2          $ \\ 
Mrk\,771            & $~~13.0 \pm 0.8            $ & ~~~~~~~~~~$    \ldots      $ & ~~~~~$0.7 \pm 0.4          $ & $ 15.8 \pm 0.8            $ & ~~~~~~~~~~$    \ldots      $ & ~~~~~$0.6 \pm 0.4          $ \\ 
NGC\,4593           & $ 108.2 \pm 8.5            $ & ~~~~~~~~~~$    \ldots      $ & ~~~~~$1.0 \pm 0.3          $ & $ 19.5 \pm 1.2            $ & ~~~~~~~~~~$    \ldots      $ & ~~~~~$1.0 \pm 0.3          $ \\ 
Mrk\,279            & $~~25.7 \pm 1.7            $ & ~~~~~$1.2 \pm 0.2^{\,\rm x}$ & ~~~~~$0.6 \pm 0.2^{\,\rm x}$ & $ 14.1 \pm 0.8            $ & ~~~~~$1.1 \pm 0.2^{\,\rm x}$ & ~~~~~$0.5 \pm 0.2^{\,\rm x}$ \\ 
NGC\,5506           & $~~92.4 \pm 1.0^{\,\rm x}  $ & ~~~~~$2.0 \pm 0.2^{\,\rm x}$ & ~~~~~$1.1 \pm 0.2^{\,\rm x}$ & $~~9.4 \pm 0.5            $ & ~~~~~$1.8 \pm 0.2^{\,\rm x}$ & ~~~~~$1.0 \pm 0.2^{\,\rm x}$ \\ 
NGC\,5548           & $~~47.3 \pm 3.3            $ & ~~~~~$1.9 \pm 0.3          $ & ~~~~~$1.3 \pm 0.3          $ & $ 16.2 \pm 1.1            $ & ~~~~~$1.8 \pm 0.3          $ & ~~~~~$1.2 \pm 0.3          $ \\ 
Ark\,479            & $~~17.0 \pm 0.9            $ & ~~~~~~~~~~$    \ldots      $ & ~~~~~$1.3 \pm 0.2^{\,\rm x}$ & $~~6.5 \pm 0.3            $ & ~~~~~~~~~~$    \ldots      $ & ~~~~~$1.2 \pm 0.2^{\,\rm x}$ \\ 
Mrk\,506            & $~~21.8 \pm 0.5            $ & ~~~~~$2.3 \pm 0.2^{\,\rm x}$ & ~~~~~$1.5 \pm 0.2^{\,\rm x}$ & $ 17.3 \pm 0.4            $ & ~~~~~$2.1 \pm 0.2^{\,\rm x}$ & ~~~~~$1.4 \pm 0.2^{\,\rm x}$ \\ 
Mrk\,507            & $~~10.1 \pm 0.9            $ & ~~~~~$2.7 \pm 0.4          $ & ~~~~~$1.6 \pm 0.3          $ & $ 10.5 \pm 0.4            $ & ~~~~~$2.4 \pm 0.4          $ & ~~~~~$1.5 \pm 0.3          $ \\ 
3C\,382             & $~~17.8 \pm 2.2            $ & ~~~~~$2.4 \pm 0.3          $ & ~~~~~$1.4 \pm 0.3          $ & $ 21.0 \pm 2.4            $ & ~~~~~$2.0 \pm 0.3          $ & ~~~~~$1.3 \pm 0.3          $ \\ 
3C\,390.3           & $~~11.6 \pm 0.6            $ & ~~~~~$2.6 \pm 0.3          $ & ~~~~~$1.4 \pm 0.2          $ & $ 14.1 \pm 0.7            $ & ~~~~~$2.2 \pm 0.3          $ & ~~~~~$1.3 \pm 0.2          $ \\ 
NGC\,6814           & $~~91.4 \pm 4.0            $ & ~~~~~$2.9 \pm 0.2          $ & ~~~~~$2.0 \pm 0.2          $ & $~~9.7 \pm 0.7            $ & ~~~~~$2.4 \pm 0.2          $ & ~~~~~$1.7 \pm 0.2          $ \\ 
Mrk\,509            & $~~16.8 \pm 0.9            $ & ~~~~~$2.6 \pm 0.2          $ & ~~~~~$1.7 \pm 0.1          $ & $ 10.8 \pm 0.4            $ & ~~~~~$2.3 \pm 0.2          $ & ~~~~~$1.5 \pm 0.1          $ \\ 
Mrk\,1513           & $~~15.3 \pm 1.0            $ & ~~~~~~~~~~$    \ldots      $ & ~~~~~$1.1 \pm 0.5          $ & $ 16.1 \pm 0.9            $ & ~~~~~~~~~~$    \ldots      $ & ~~~~~$1.0 \pm 0.5          $ \\ 
Mrk\,304            & $~~11.4 \pm 0.5            $ & ~~~~~$2.0 \pm 0.2          $ & ~~~~~$1.3 \pm 0.2          $ & $ 15.4 \pm 0.6^{\,\rm x}  $ & ~~~~~$1.6 \pm 0.2          $ & ~~~~~$1.1 \pm 0.2          $ \\ 
Ark\,564            & $~~23.7 \pm 0.2^{\,\rm x}  $ & ~~~~~$1.4 \pm 0.3^{\,\rm x}$ & ~~~~~$1.0 \pm 0.3^{\,\rm x}$ & $ 10.9 \pm 0.1^{\,\rm x}  $ & ~~~~~$1.2 \pm 0.3^{\,\rm x}$ & ~~~~~$0.9 \pm 0.3^{\,\rm x}$ \\ 
NGC\,7469           & $~~49.4 \pm 1.1            $ & ~~~~~$1.9 \pm 0.2          $ & ~~~~~$1.2 \pm 0.1          $ & $ 14.5 \pm 0.3            $ & ~~~~~$1.7 \pm 0.2          $ & ~~~~~$1.1 \pm 0.1          $ \\ 
Mrk\,315$^{\,\rm b}$& $~~14.5 \pm 0.8            $ & ~~~~~$2.1 \pm 0.6          $ & ~~~~~$1.0 \pm 0.4          $ & $ 12.9 \pm 1.8            $ & ~~~~~$1.5 \pm 0.6          $ & ~~~~~$0.7 \pm 0.4          $ \\ 
NGC\,7603           & $~~41.1 \pm 0.2            $ & ~~~~~~~~~~$    \ldots      $ & ~~~~~$1.1 \pm 0.3          $ & $ 21.6 \pm 1.2            $ & ~~~~~~~~~~$    \ldots      $ & ~~~~~$1.0 \pm 0.3          $ \\ 
Mrk\,541            & $~~18.9 \pm 2.3            $ & ~~~~~~~~~~$    \ldots      $ & ~~~~~$1.1 \pm 0.2          $ & $ 12.3 \pm 1.7            $ & ~~~~~~~~~~$    \ldots      $ & ~~~~~$0.9 \pm 0.2          $ \\ 
\noalign{\smallskip}
\noalign{\smallskip}
Med./MAD        & $~~~~~~~~\ldots               $ & ~~~~~$2.2\,\,/\,\,0.4            $ & ~~~~~$1.3\,\,/\,\,0.2            $ & $13.9\,\,/\,\,3.7               $ & ~~~~~$1.9\,\,/\,\,0.3            $ & ~~~~~$1.1\,\,/\,\,0.3            $ \\

\hline
\end{tabular}
\end{center}

$^{\rm a}$ The 25 $B$ mag arcsec$^{-2}$ isophote was used. \\
$^{\rm b}$ Chatzichristou's (\cite{C1_00}) images available through NED were used. \\
$^{\rm x}$ The SMA or CI were extrapolated; the extrapolated SMA and CI were not taken into account in the weight-averaging of multi-epoch data unless all the data are extrapolated.
\end{table*}

\section{Bar parameters} 
\label{barpar}

We consider a galaxy barred if there is an ellipticity maximum greater than 0.16 with an amplitude of at least 0.08 over a region of PA constant within $20\degr$, following Aguerri et~al. (\cite{AMC_09}). At the transition to the disk the PA changes unless the bar and disk are aligned. 

The ellipticity maximum and the corresponding PA in the region of the bar are adopted as its ellipticity ($\epsilon_{\,\rm bar}$) and PA ($PA_{\rm bar}$), respectively. The presence of a bulge, together with eventual boxiness of the bar isophotes, leads to underestimation of the bar ellipticity, obtained on the base of ellipse fits, especially in galaxies with big bulges (see also Men\'endez-Delmestre et~al. \cite{MSS_07}). Generally, the PA along bars is well constrained thanks to the $x_1$ orbits (e.g., Athanassoula \cite{A_92b}).

As bar length, $\ell$, we adopted the SMA, where the ellipticity decreases with 15\% from its maximal value, after Martinez-Valpuesta, Shlosman \& Heller (\cite{MSH_06}), according to whom the so estimated length is consistent with the size of the maximal stable $x_1$ orbit; bar length has already been derived in that manner by Fathi et~al. (\cite{FBP_09}). The post-maximum ellipticity slope, generally steeper than the pre-maximum one, is often influenced by spiral arm beginnings or rings. To reduce this influence, we took the minimum of the SMAs, corresponding to the 15\% ellipticity decrease, both before and after the ellipticity maximum. 
The error of $\ell$ and $\ell_{\rm max}$ was estimated as the SMA change relevant to $1\,\sigma$ change of the  corresponding ellipticity. 

This approach worked well for all barred galaxies\footnote{We consider only large-scale bars. Given also in Table\,\ref{T_barpar} are the parameters of the nuclear bar of Mrk\,352, firstly reported in Paper\,I, but they were not taken into account in any of the correlations below, nor in the estimation of the median parameters.} but NGC\,3227, NGC\,4051, and NGC\,4593, for which it overestimates the bar lengths. Not taking these galaxies into account, we found $\ell$ and $\ell_{\rm max}$ to correlate tightly (with a Pearson correlation coefficient of 0.999 and a probability that this is achieved by uncorrelated points below $10^{-6}$, Fig.\,\ref{B_LmaxLa_lgr}). The median value of the ratio $\ell/\ell_{\rm max}$ is 1.22 with a MAD of 0.06. The bar lengths of the above three galaxies were estimated using the corresponding $\ell_{\rm max}$ values and the median $\ell/\ell_{\rm max}$ ratio.

We used the $I_{\rm \scriptstyle C}$ profiles to estimate the bar parameters as bars are best pronounced and the spiral structure influence is minimized there. The derived parameters are listed in Table\,\ref{T_barpar}; multi-epoch bar parameters were weight-averaged except for Mrk\,79\footnote{Only the data of better seeing were taken into account (see Paper\,I).}. The bar parameters of Mrk\,352, Mrk\,771, and Mrk\,279 were estimated using Hubble Space Telescope (HST) data and the ones of NGC\,6814~-- using Two Micron All Sky Survey (2MASS) data (see Paper\,I). 

The bar length was deprojected following Martin (\cite{M_95}):
\begin{displaymath}
\ell^{\,(i)}=\ell\sqrt{\cos^2\Theta+\sec^2i\,\sin^2\Theta\,},\nonumber
\end{displaymath}
where $\Theta$ is the angle between the SMAs of the bar and disk. To compute the deprojected bar ellipticity, $\epsilon_{\;\rm bar}^{\,(i)}$, the bar semi-minor axis was deprojected using the above formula, multiplied by 1\,--\,$\epsilon_{\,\rm bar}$; now $\Theta$ is the angle between the semi-minor axis of the bar and the SMA of the disk. 

The median of the deprojected bar ellipticity, length, and relative length, $\ell^{\,(i)}/a_{24}^{(0)}$, are 0.39\,(0.12), 5.44\,(1.80)\,kpc, and 0.44\,(0.11), respectively, with the corresponding MADs given in parentheses. 

Figure\,\ref{B_AvLi} shows $\ell^{\,(i)}$ as a function of $a_{24}^{(0)}$, the Pearson correlation coefficient is 0.783 (with a probability that this is achieved by uncorrelated points about $\,10^{-4}$); a weaker correlation was found by Laine et~al. (\cite{LSK_02}). 

Deprojected bar ellipticity can be used as a first-order approximation of bar strength (e.g., Laurikainen et~al. \cite{LSR_02}; Block et~al. \cite{BBK_04}). 
We classified a bar as strong if $\epsilon_{\;\rm bar}^{\,(i)}$$\,>\,$$0.45$ after Laine et~al. (\cite{LSK_02}). 
This resulted in three strong bars among the 17 barred galaxies. In particular, Seyfert bars appear weaker than their inactive counterparts at the 95\% confidence level (see Paper\,I).
The deprojected bar ellipticities of the matched inactive sample used in this comparison are listed in Appendix\,\ref{AppendixB}.
Concerning the morphological classification presented in Paper\,I, the barred galaxies of both samples were given designations ``AB'' or ``B'' according to the values of their deprojected ellipticities.

The bar-like structures of Mrk\,595, Mrk\,279, and NGC 7469 are most probably ovals/lenses, given their deprojected ellipticities below 0.15 (\cite{KK_04});  to further specify this, kinematic data are needed though\linebreak (\cite{SW_93}). Note that bars, ovals, and\linebreak lenses are essentially equivalent regarding gas inflow\linebreak (\cite{KK_04}).

The deprojected relative bar length and bar ellipticity show no clear correlation (Fig.\,\ref{B_EiLni}; see also Fig.\,5 of M\'arquez et~al. \cite{MDM_00} and Fig.\,6 of Laine et~al. \cite{LSK_02}). Overplotted are the empirical limits of long ($\ell^{\,(i)}/a_{24}^{(0)}$$\,\geq\,$$0.18$) and strong ($\epsilon$$\,\geq\,$$0.4$) bars of Martinet \& Friedli (\cite{MF_97})\footnote{Note that our definition of bar length differs from that used by the authors.}, dotted, and of strong bars ($\epsilon$$\,>\,$$0.45$) of Laine et~al. (\cite{LSK_02}), dashed.

\begin{table*}[t]
\caption{Bar parameters, estimated in $I_{\rm \scriptstyle C}$, except for the galaxies, for which HST or 2MASS data were used.}
\label{T_barpar}
\begin{center}
\begin{tabular}{@{}l@{\hspace{0.3cm}}rllrrr@{\hspace{0.3cm}}rl@{}}
\hline
\noalign{\smallskip}
~~~Galaxy & $PA_{\rm bar}$~~ & ~~~~~~~~~$\epsilon_{\,\rm bar}$ & ~~~~~~~$\epsilon_{\,\rm bar}^{\,(i)}$ & $\ell_{\rm max}$\,~~~ & $\ell$\,~~~~~~~ & \multicolumn{2}{c}{$\ell^{\,(i)}$} & ~~~$\ell^{\,(i)}/a_{24}^{(0)}$ \\
\cline{7-8}
\noalign{\smallskip}
       & (degree)~ &     &     & (arcsec)~  & (arcsec)~~ & (arcsec)~ & (kpc)\,~~~~ &   \\
\noalign{\smallskip}\hline
\noalign{\smallskip}

Mrk\,352$^{\rm a,b}$& $176.9 \pm 2.0$ & $0.158 \pm 0.010$ & $0.34 \pm 0.02$ & $ 2.0 \pm 0.2$ & $ 2.2 \pm 0.1$ & $ 2.8 \pm 0.1$ & $ 0.75 \pm 0.03$ & $0.18 \pm 0.01$ \\
Mrk\,573            & $178.5 \pm 1.7$ & $0.339 \pm 0.016$ & $0.42 \pm 0.01$ & $ 9.1 \pm 0.6$ & $10.2 \pm 0.1$ & $11.8 \pm 0.3$ & $ 3.73 \pm 0.10$ & $0.36 \pm 0.02$ \\
Mrk\,595            & $111.2 \pm 0.7$ & $0.365 \pm 0.008$ & $0.13 \pm 0.01$ & $ 6.5 \pm 0.9$ & $ 8.5 \pm 0.2$ & $ 9.0 \pm 0.3$ & $ 4.57 \pm 0.13$ & $0.43 \pm 0.02$ \\
Mrk\,376            & $  9.9 \pm 0.3$ & $0.511 \pm 0.005$ & $0.39 \pm 0.03$ & $ 4.3 \pm 0.3$ & $ 5.3 \pm 0.1$ & $ 5.8 \pm 0.1$ & $ 6.06 \pm 0.12$ & $0.41 \pm 0.04$ \\
Mrk\,79$^{\rm c}$   & $ 57.1 \pm 0.5$ & $0.596 \pm 0.007$ & $0.66 \pm 0.03$ & $14.9 \pm 1.0$ & $19.4 \pm 0.2$ & $22.7 \pm 0.6$ & $ 9.94 \pm 0.28$ & $0.55 \pm 0.05$ \\
Mrk\,382            & $  5.6 \pm 1.3$ & $0.390 \pm 0.030$ & $0.33 \pm 0.03$ & $ 6.7 \pm 0.3$ & $ 8.0 \pm 0.3$ & $ 8.4 \pm 0.4$ & $ 5.53 \pm 0.25$ & $0.40 \pm 0.02$ \\
NGC\,3227           & $150.0 \pm 2.2$ & $0.603 \pm 0.032$ & $0.21 \pm 0.08$ & $51.9 \pm 6.8$ & $63.3 \pm 8.9$ & $63.3 \pm 8.9$ & $ 6.21 \pm 0.87$ & $0.57 \pm 0.08$ \\
NGC\,3516           & $167.8 \pm 0.1$ & $0.279 \pm 0.001$ & $0.34 \pm 0.01$ & $11.8 \pm 0.3$ & $14.1 \pm 0.1$ & $16.6 \pm 0.3$ & $ 2.98 \pm 0.05$ & $0.29 \pm 0.02$ \\
NGC\,4051           & $134.6 \pm 0.7$ & $0.600 \pm 0.010$ & $0.45 \pm 0.11$ & $60.9 \pm 2.9$ & $74.3 \pm 5.1$ & $86.2 \pm 8.1$ & $ 5.26 \pm 0.49$ & $0.53 \pm 0.05$ \\
NGC\,4151           & $128.2 \pm 0.8$ & $0.510 \pm 0.010$ & $0.54 \pm 0.02$ & $65.3 \pm 3.7$ & $76.5 \pm 0.6$ & $81.7 \pm 2.7$ & $ 6.70 \pm 0.22$ & $0.81 \pm 0.03$ \\
Mrk\,766            & $106.0 \pm 0.7$ & $0.457 \pm 0.009$ & $0.43 \pm 0.03$ & $10.1 \pm 1.2$ & $13.3 \pm 0.2$ & $14.7 \pm 0.4$ & $ 3.99 \pm 0.12$ & $0.50 \pm 0.02$ \\
Mrk\,771$^{\rm a}$  & $ 30.9 \pm 1.3$ & $0.429 \pm 0.015$ & $0.44 \pm 0.04$ & $ 5.0 \pm 0.4$ & $ 6.0 \pm 0.2$ & $ 6.6 \pm 0.3$ & $ 7.77 \pm 0.31$ & $0.49 \pm 0.03$ \\
NGC\,4593           & $ 56.7 \pm 0.2$ & $0.627 \pm 0.004$ & $0.44 \pm 0.14$ & $47.6 \pm 4.2$ & $58.1 \pm 5.9$ & $60.6 \pm 6.5$ & $12.11 \pm 1.30$ & $0.62 \pm 0.08$ \\
Mrk\,279$^{\rm a}$  & $ 31.1 \pm 0.4$ & $0.430 \pm 0.004$ & $0.13 \pm 0.01$ & $ 5.6 \pm 0.4$ & $ 7.3 \pm 0.3$ & $ 7.3 \pm 0.3$ & $ 4.29 \pm 0.15$ & $0.30 \pm 0.02$ \\
Ark\,479            & $105.3 \pm 0.1$ & $0.474 \pm 0.001$ & $0.25 \pm 0.02$ & $ 5.6 \pm 0.2$ & $ 7.2 \pm 0.1$ & $ 7.3 \pm 0.1$ & $ 2.87 \pm 0.02$ & $0.44 \pm 0.02$ \\
NGC\,6814$^{\rm d}$ & $ 28.0 \pm 0.8$ & $0.288 \pm 0.006$ & $0.33 \pm 0.01$ & $12.1 \pm 1.3$ & $15.8 \pm 0.1$ & $17.0 \pm 0.4$ & $ 1.51 \pm 0.04$ & $0.16 \pm 0.01$ \\
Ark\,564            & $ 37.8 \pm 1.5$ & $0.460 \pm 0.020$ & $0.57 \pm 0.01$ & $ 8.1 \pm 0.7$ & $ 9.3 \pm 0.1$ & $11.9 \pm 0.2$ & $ 5.44 \pm 0.09$ & $0.50 \pm 0.01$ \\
NGC\,7469           & $120.4 \pm 0.7$ & $0.360 \pm 0.010$ & $0.12 \pm 0.01$ & $16.0 \pm 0.4$ & $18.5 \pm 0.2$ & $18.5 \pm 0.2$ & $ 5.47 \pm 0.07$ & $0.38 \pm 0.01$ \\
\noalign{\smallskip}
\noalign{\smallskip}
Med./MAD          &      \ldots~~~~~  &        ~~~~~~~~~\,\ldots    & 0.39~~/~~0.12    &\ldots~~~~~~&       \ldots~~~~~~&       \ldots~~~~~  & 5.44~~/~~1.80     & 0.44~~/~~\,0.11      \\
\hline
\end{tabular}
\end{center}
$^{\rm a}$ The bar parameters were estimated using archival HST data. \\
$^{\rm b}$ Not taken into account in the computation of the median parameters (see text). \\
$^{\rm c}$ The multi-epoch bar parameters were not weight-averaged. \\
$^{\rm d}$ The bar parameters were estimated using 2MASS data.
\end{table*}

\section{Summary}
\label{summary}
This paper is third in a series, studying the optical properties of a sample of Seyfert galaxies.
The first paper addresses the evidence of non-axisymmetric perturbation of the potential in a sample of 35 Seyfert galaxies and in a matched inactive sample.
A homogeneous set of global (ellipticity, PA, inclination, and total magnitude) and isophotal (SMA and CIs at 24 $V$ mag arcsec$^{-2}$) parameters of the Seyfert sample are reported in this study. 
Correction for galactic absorption, inclination and internal absorption, cosmological dimming, as well as $K$- and $E$-correction was applied to the isophotal parameters.
We found the following median isophotal parameters:
\begin{eqnarray}
a_{24}^{\rm (0)}&=&13.9\,\rm kpc, \nonumber \\
(B-I_{\rm \scriptstyle C})_{24}&=&2.2\,\rm mag\,arcsec^{-2}, \nonumber \\
(V-I_{\rm \scriptstyle C})_{24}&=&1.3\,\rm mag\,arcsec^{-2}, \nonumber \\
(B-I_{\rm \scriptstyle C})_{24}^{(0)}&=&1.9\,\rm mag\,arcsec^{-2}, \nonumber \\
(V-I_{\rm \scriptstyle C})_{24}^{(0)}&=&1.1\,\rm mag\,arcsec^{-2}. \nonumber
\end{eqnarray}
The estimated parameters can be further used in various galactic structure studies.

We presented a set of bar parameters~-- ellipticity, PA, SMA corresponding to the ellipticity maximum in the bar region, and length; deprojected values of the bar ellipticity, length, and relative length in terms of galaxy isophotal SMA are also given.
As bar length we adopted the minimum of the SMAs corresponding to 15\% ellipticity decrease from its maximal value, both before and after the ellipticity maximum. 
The so obtained bar length and the most often used bar length estimate~-- the SMA, corresponding to the ellipticity maximum, show a tight correlation with a median ratio $\ell/\ell_{\rm max}$ of 1.22, which we further used to obtain the bar length in cases the above approach did not work. 
The median of the deprojected bar ellipticity, length, and relative length are 0.39, 5.44\,kpc, and 0.44, respectively. 
The deprojected bar length correlates with the corrected isophotal SMA at 24 $V$ mag arcsec$^{-2}$. 
Seventeen of the galaxies have large-scale bars, three of which are strong, based on the deprojected bar ellipticity as a rough estimate of bar strength. 
The deprojected relative bar length and  bar ellipticity show no clear correlation.

Global ellipticities and deprojected bar ellipticities of the matched inactive sample are also presented.

\acknowledgements
We thank the anonymous referee for the useful recommendations.

We thank Dr. P. Nedyalkov for the useful discussion regarding dust extinction.

This research has made use of the NASA/IPAC Extragalactic Database (NED) which is operated by the Jet Propulsion Laboratory, California Institute of Technology, under contract with the National Aeronautics and Space Administration.

We acknowledge the usage of the HyperLeda database\linebreak (http://leda.univ-lyon1.fr).

Some of the data presented in this paper were obtained from the Multimission Archive at the Space Telescope Science
Institute (MAST). STScI is operated by the Association of Universities for Research in Astronomy, Inc., under NASA
contract NAS5-26555. Support for MAST for non-HST data is provided by the NASA Office of Space Science via grant
NAG5-7584 and by other grants and contracts.

This publication makes use of data products from the Two Micron All Sky Survey, which is a joint project of the
University of Massachusetts and the Infrared Processing and Analysis Center/California Institute of Technology, funded by
the National Aeronautics and Space Administration and the National Science Foundation.

\begin{appendix}
\section{Details about the galaxies Mrk\,1040, NGC\,5506, and Mrk\,507}
\label{AppendixA}
\begin{figure*}[htbp]
   \centering
\includegraphics[width=5.4cm]{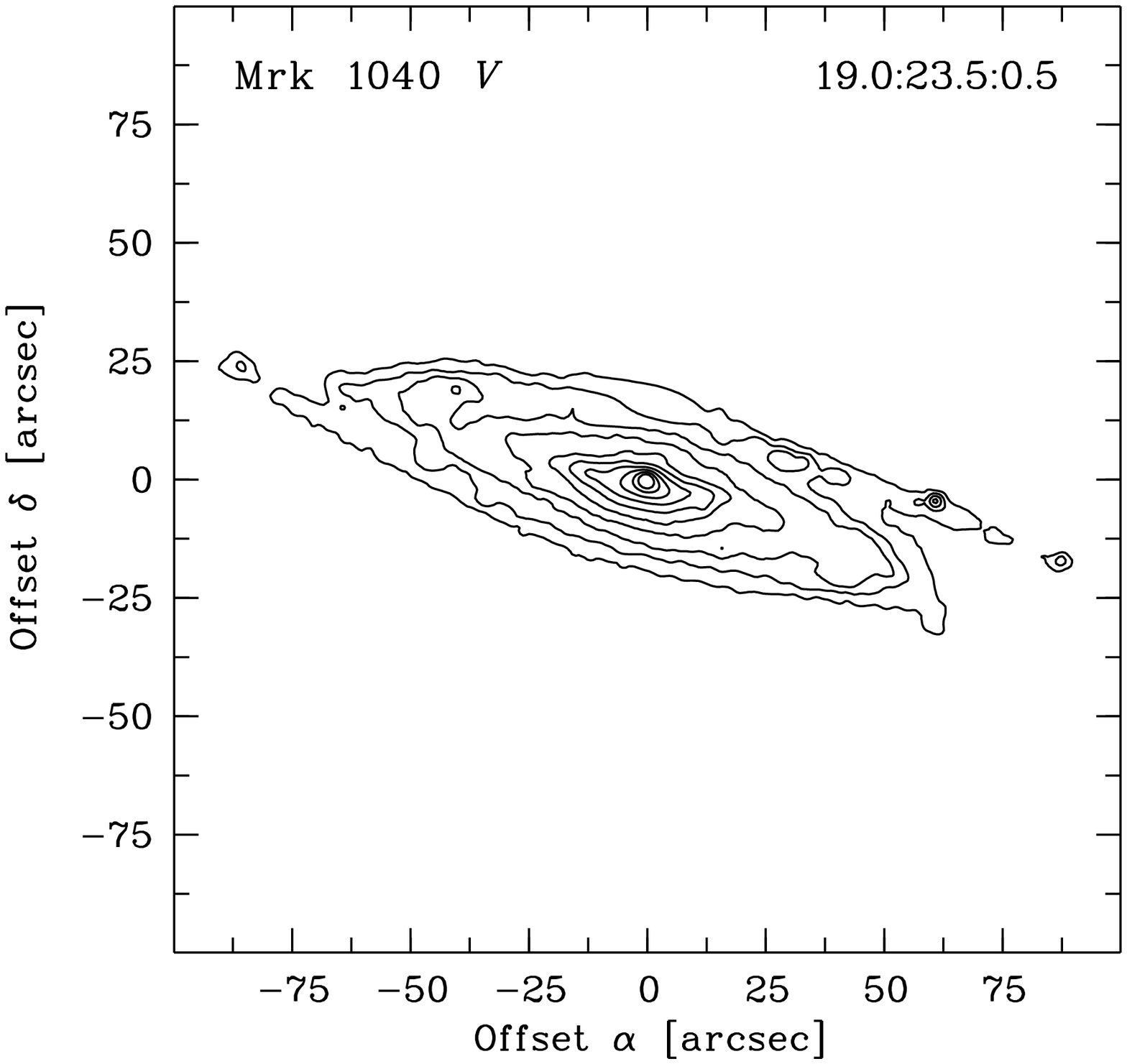}
\hspace{0.2cm}
\includegraphics[width=5.4cm]{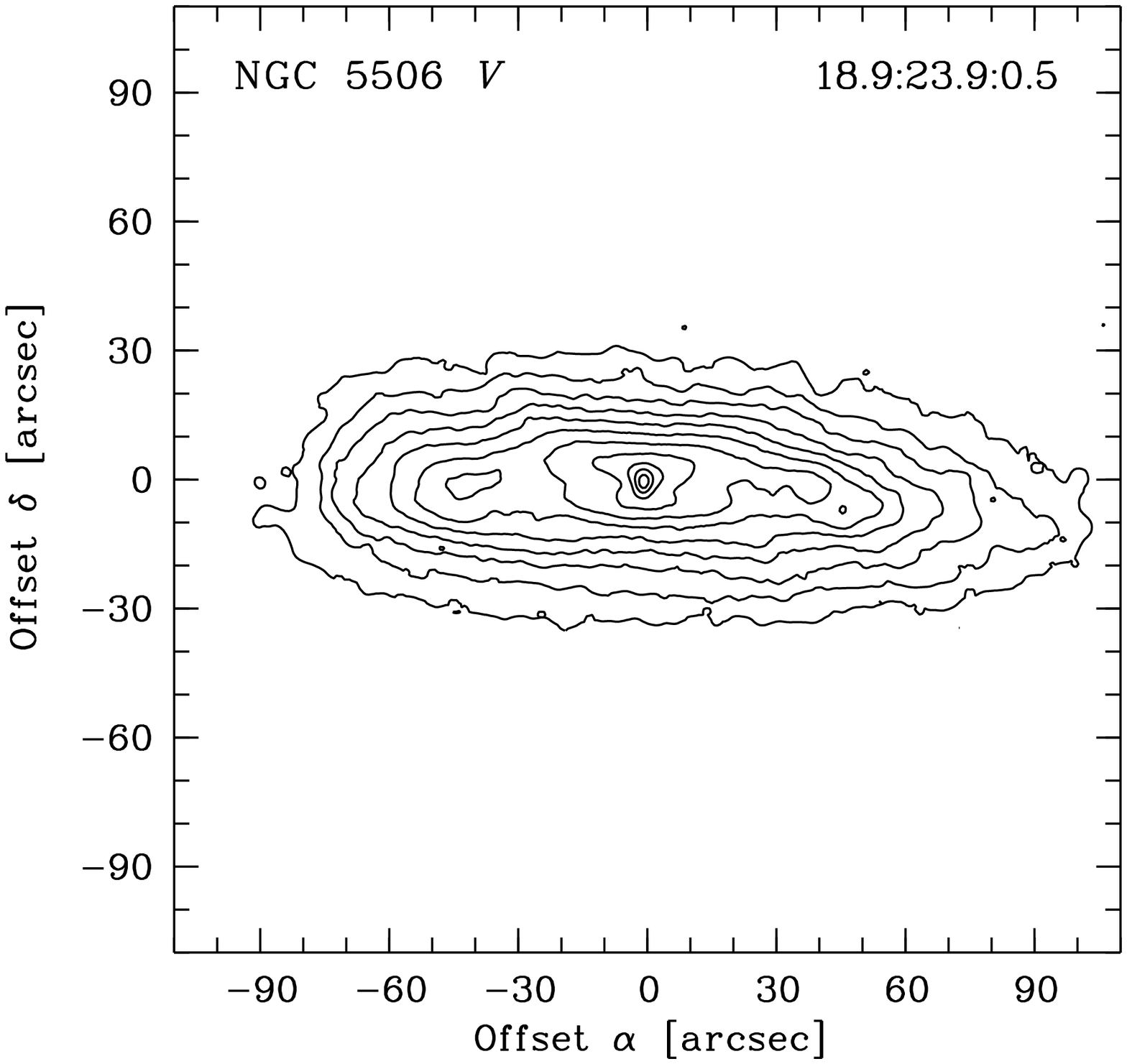}
\hspace{0.2cm}
\includegraphics[width=5.4cm]{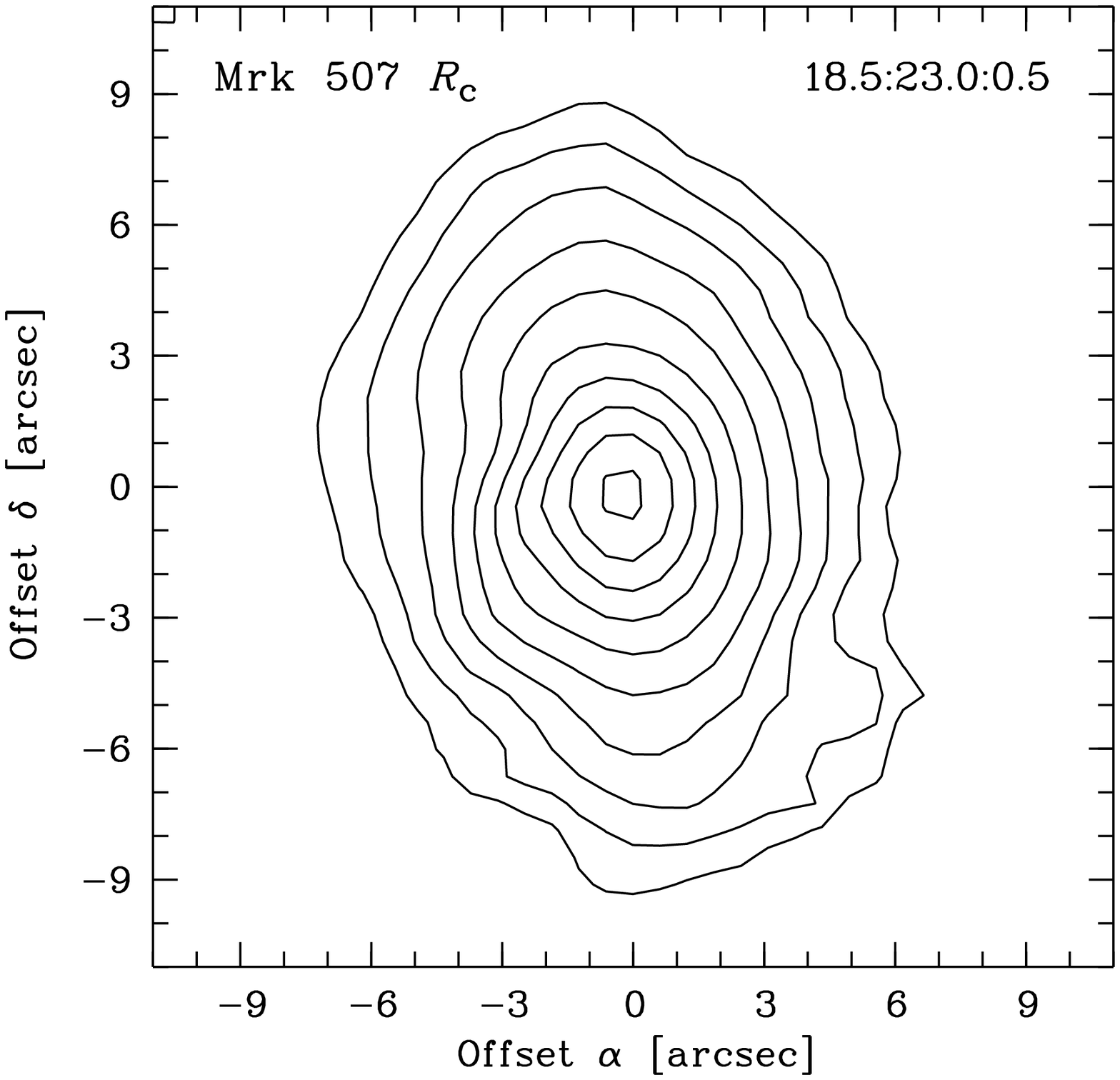}

\vspace{0.3cm}

\includegraphics[width=5.4cm]{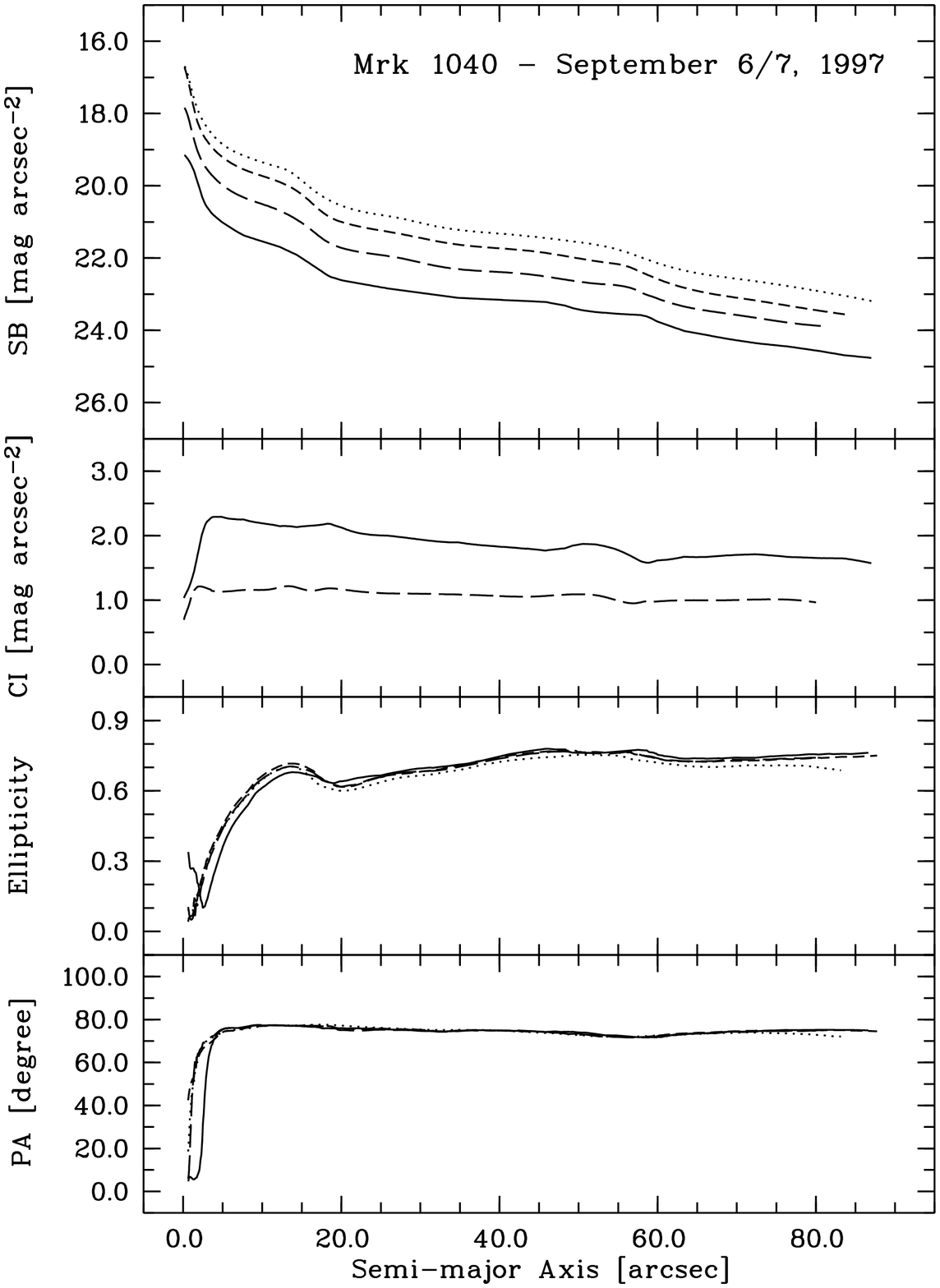}
\hspace{0.2cm}
\includegraphics[width=5.4cm]{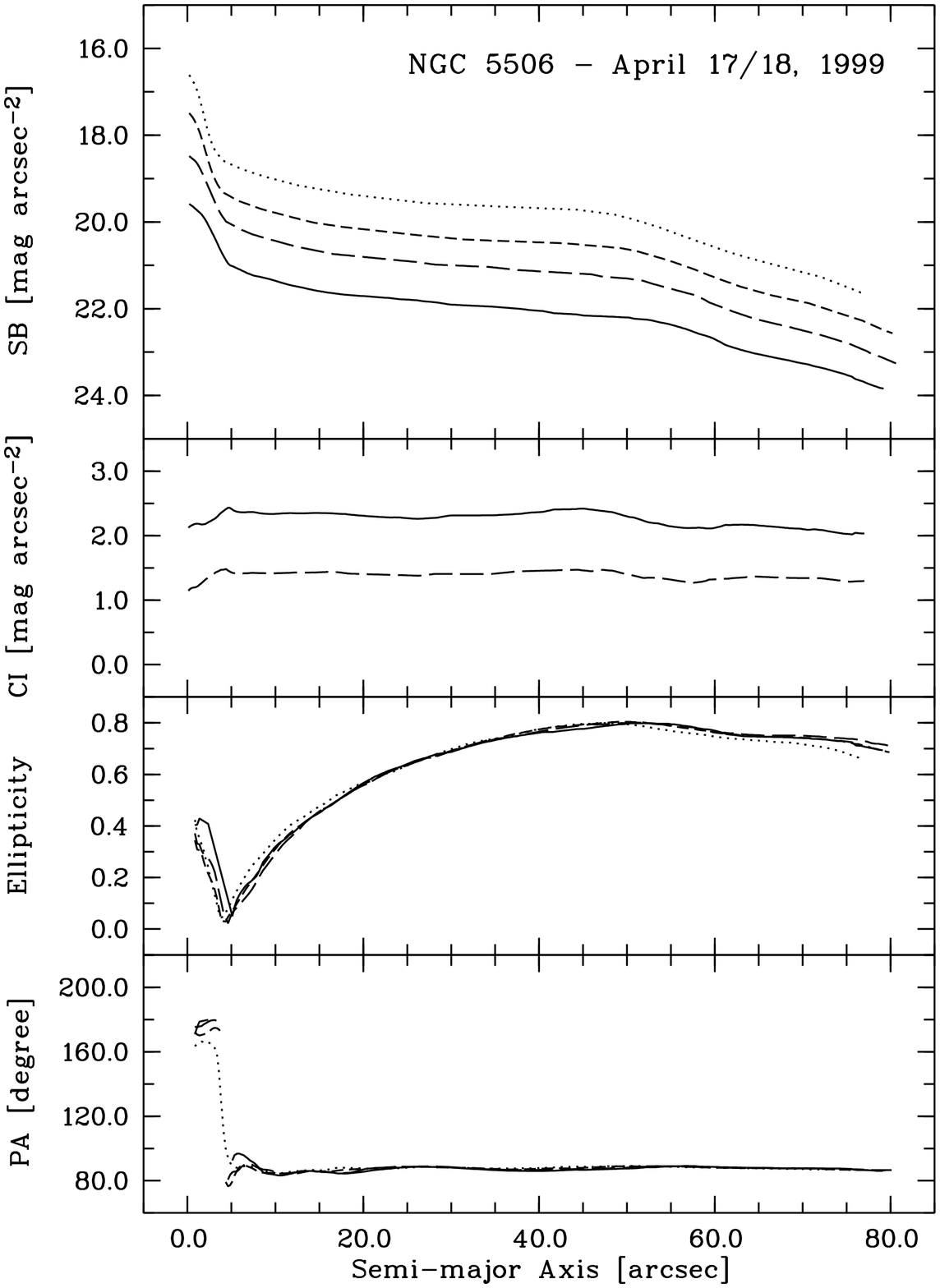}
\hspace{0.2cm}
\includegraphics[width=5.4cm]{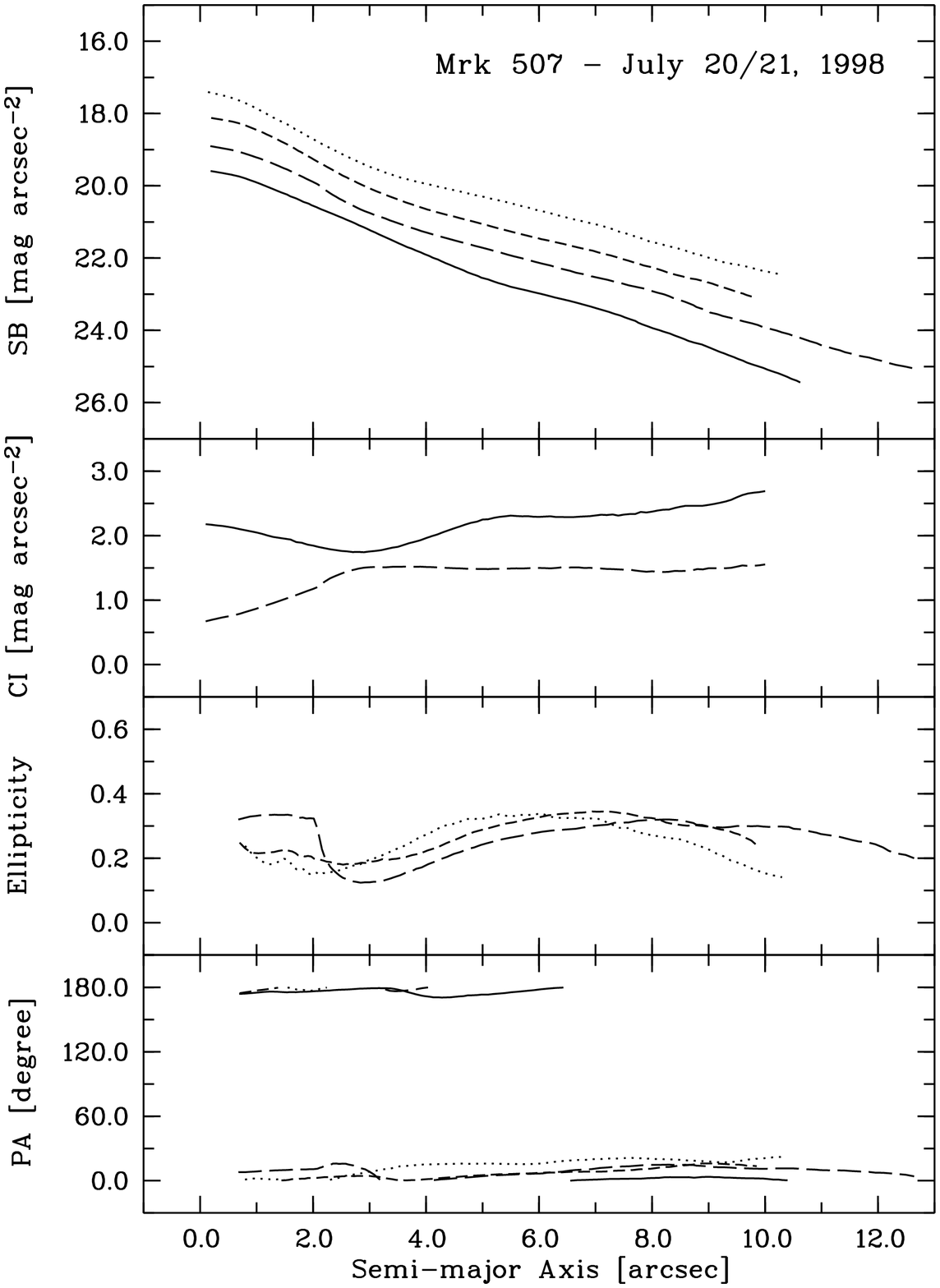}
\caption{Calibrated contour maps and profiles of the extra added galaxies, ordered by right ascension. Upper panels: contour maps. North is up and east to the left. The galaxy names and passbands are denoted in the upper left; the numbers in the upper right denote the start SB, end SB, and SB step in units of mag\,arcsec$^{-2}$. Lower panels: Profiles of the SB, CI, ellipticity, and PA. The CI profiles shown are $B\,$--$\,I_{\rm \scriptstyle C}$ (solid) and  $V\,$--$\,I_{\rm \scriptstyle C}$ (dashed). For the rest of the profiles the solid, long-dashed, short-dashed, and dotted line is for the $B$-, $V$-, $R_{\rm \scriptstyle C}$-, and $I_{\rm \scriptstyle C}$-band, respectively.}
\label{contprof}
   \end{figure*}

The galaxies were observed at the Rozhen National Astronomical Observatory, Bulgaria, with the 2-m Ritchey-Chr\'etien telescope equipped with $1024\,$$\times$$\,1024$ Photometrics AT200 CCD camera (CCD chip SITe~SI003AB with a square pixel size of $24\,\mu$m that corresponds to $0\farcs309$ on the sky). Standard Johnson-Cousins $BVR_{\rm \scriptstyle C}I_{\rm \scriptstyle C}$ filters were used. We present details about the observations, image quality, and standard fields used for calibration of the extra added galaxies in Table\,\ref{T_obs}. 
Calibrated contour maps and profiles of the SB, CI, ellipticity, and PA are given in Fig.\,\ref{contprof}.
Regarding Mrk\,507, a stellar-like object (\cite{HO_87};\linebreak \cite{GHM_03}) of comparable brightness is located about $2\arcsec$ from its nucleus at $\rm PA\,$$=$$\,110\degr$.
We cleaned the projected object with the help of Moffat PSF subtraction; we should, however, keep in mind the close proximity of the object and the small size of the galaxy itself.

\begin{table*}
\caption{Log of the observations.}
\label{T_obs}
\begin{center}
\begin{tabular}{@{}lllll@{}}
\hline
\noalign{\smallskip}
~~Galaxy & ~~Civil Date & ~~~FWHM & ~~~~~~~$\beta$ & Calibration  \\
       & (yyyy mm dd) & ~~~(arcsec) &    &     \\
\noalign{\smallskip}
~~~~~(1) & ~~~~~~~(2) & ~~~~~~~(3) & ~~~~~~~(4) & ~~~~~~(5) \\
\noalign{\smallskip}
\hline
\noalign{\smallskip}

Mrk\,1040          & 1997 09 06 & $1.36 \pm 0.02$ & $2.98 \pm 0.19$ & SS   \\
NGC\,5506          & 1999 04 17 & $2.51 \pm 0.04$ & $~~~~~~\,\ldots$& M\,92 \\
Mrk\,507           & 1998 07 20 & $1.96 \pm 0.02$ & $2.45 \pm 0.05$ & M\,92\\
\hline
\end{tabular}
\end{center}
Key to columns. (3) and (4) FWHM (full width at half maximum) and power-law index ($\beta$), respectively, of the Moffat PSF (point spread function); ellipsis dots denote Gaussian PSF was assumed; (5) Standard fields used for calibration: SS~-- secondary standards used.
\end{table*}

\section{Global ellipticities and deprojected bar ellipticities of the matched inactive galaxies}
\label{AppendixB}
We present the global ellipticities of the inactive galaxies and the deprojected bar ellipticities of the barred subsample in Table\,\ref{T_inactive}. Both sets of ellipticities were estimated as for the Sy sample.
The median of the deprojected bar ellipticity is 0.49 with a MAD of 0.14.
The galaxy ellipticities were used in the sample matching, and the bar ellipticities were involved in the bar strength comparison of the two samples (see Paper\,I).

\begin{table}[t]
\begin{center}
\caption{Global ellipticities, estimated over all available passbands, and deprojected bar ellipticities, determined in the reddest available passband, of the matched inactive galaxies.}
\label{T_inactive}
\begin{tabular}{@{}llcc@{}}
\hline
\noalign{\smallskip}
Seyfert & Inactive & $\epsilon$ & $\epsilon^{(i)}_{\,\rm bar}$ \\
\noalign{\smallskip}
\hline
\noalign{\smallskip}
Mrk 335  & IC 5017                  & 0.18 &   0.55 \\
III Zw 2 & 2MASX J01505708+0014040  & 0.30 & \ldots \\
Mrk 348  & NGC 2144                 & 0.16 & \ldots \\
I Zw 1   & ESO 155-- G 027          & 0.07 &   0.68 \\
Mrk 352  & 2MASX J04363658--0250350 & 0.26 & \ldots \\
Mrk 573  & ESO 542-- G 015          & 0.08 &   0.29 \\
Mrk 590  & NGC 4186                 & 0.20 & \ldots \\
Mrk 595  & 2MASX J00342513--0735582 & 0.42 &   0.14 \\
3C 120   & ESO 202-- G 001          & 0.39 & \ldots \\
Ark 120  & IC 5065                  & 0.10 &   0.12 \\
Mrk 376  & ESO 545-- G 036          & 0.12 & \ldots \\
Mrk 79   & ESO 340-- G 036          & 0.13 &   0.51 \\
Mrk 382  & ESO 268-- G 032          & 0.26 &   0.44 \\
NGC 3227 & IC 5240                  & 0.27 &   0.64 \\
NGC 3516 & ESO 183-- G 030          & 0.20 & \ldots \\
NGC 4051 & IC 1993                  & 0.05 & \ldots \\
NGC 4151 & NGC 2775                 & 0.20 & \ldots \\
Mrk 766  & UGC 6520                 & 0.32 &   0.47 \\
Mrk 771  & ESO 349-- G 011          & 0.26 &   0.56 \\
NGC 4593 & NGC 4902                 & 0.07 &   0.58 \\
Mrk 279  & ESO 324-- G 003          & 0.31 & \ldots \\
NGC 5548 & NGC 466                  & 0.12 & \ldots \\
Ark 479  & ESO 297-- G 027          & 0.42 & \ldots \\
Mrk 506  & ESO 510-- G 048          & 0.38 & \ldots \\
3C 382   & ESO 292-- G 022          & 0.34 & \ldots \\
3C 390.3 & ESO 249-- G 009          & 0.15 &   0.37 \\
NGC 6814 & NGC 7421                 & 0.20 &   0.57 \\
Mrk 509  & ESO 147-- G 013          & 0.15 & \ldots \\
Mrk 1513 & 2MASX J14595983+2046121  & 0.61 & \ldots \\
Mrk 304  & ESO 292-- G 007          & 0.36 & \ldots \\
Ark 564  & ESO 552-- G 053          & 0.14 &   0.69 \\
NGC 7469 & NGC 897                  & 0.32 & \ldots \\
Mrk 315  & ESO 423-- G 016          & 0.16 &   0.36 \\
NGC 7603 & ESO 113-- G 050          & 0.36 & \ldots \\
Mrk 541  & UGC 9532 NED04           & 0.29 &   0.40 \\
\hline
\end{tabular}
\end{center}
\end{table}
\end{appendix}
\end{document}